\documentclass[aps, pra, superscriptaddress, nofootinbib, american, reprint, floatfix]{revtex4-2}
 
\usepackage[T1]{fontenc}
\usepackage[utf8]{inputenc}
\usepackage{babel}
\usepackage{comment}
\usepackage{cancel}

\usepackage{amsmath,amssymb,esint}
\usepackage{physics} 
\usepackage{mathtools}
\usepackage{bm}
\mathtoolsset{showonlyrefs} 

\usepackage{graphicx}
\usepackage{float}    
\usepackage{placeins} 
\usepackage{subfigure}

\usepackage{xspace}
\usepackage[pagewise]{lineno} 

\newcommand{\beq}{\begin{equation}}
\newcommand{\eeq}{\end{equation}}

\usepackage{color}
\usepackage{soul}

\definecolor{red}{rgb}{1.0,0.0,0.0}
\definecolor{blu}{rgb}{0.0,0.0,1.0}
\definecolor{gre}{rgb}{0.0,1.0,0.0}

\usepackage[colorlinks=true, urlcolor=blue, citecolor=blue]{hyperref}

\begin{document}

\title{Screened Thin-Target Bremsstrahlung with Partially-Ionized High-Z Species}

\author{S.~Guinchard}
\affiliation{Ecole Polytechnique Fédérale de Lausanne (EPFL), Swiss Plasma Center (SPC), CH-1015, Lausanne, Switzerland}

\author{Y.~Savoye-Peysson}
\affiliation{CEA, IRFM, F-13108, Saint-Paul-lez-Durance, France}

\author{J.~Decker}
\affiliation{Ecole Polytechnique Fédérale de Lausanne (EPFL), Swiss Plasma Center (SPC), CH-1015, Lausanne, Switzerland}

\begin{abstract}
Bremsstrahlung emission remains a cornerstone process in the characterization of electron dynamics in diverse high-energy environments. In particular, the accurate description of thin-target electron-ion bremsstrahlung in the presence of high-$Z$ species requires careful treatment of atomic screening effects, especially when atoms are partially ionized. We present a fully analytic screening model based on a multi-Yukawa representation of the atomic potential, enabling the calculation of bremsstrahlung cross sections for arbitrary nuclear charge and ionization state, and electron energies up to a few tens of MeV. This framework extends prior treatments of neutral atoms to include partially ionized high-$Z$ elements in a fully analytic framework.
\end{abstract}

\maketitle

\noindent\textbf{Keywords:} Bremsstrahlung, Born approximation, Cross section, QED, Relativistic scattering, Screened bremsstrahlung, Furry--Sommerfeld--Maue.

\section{Introduction} \label{sec:Introduction}
Accurate and computationally efficient descriptions of thin-target bremsstrahlung remain a central objective of theoretical and applied radiation physics. Building on the foundational analyses of Sauter~\cite{Sauter34}, Bethe--Heitler~\cite{bet34}, and Racah~\cite{Racah}, together with practical prescriptions such as the so-called additivity rule~\cite{ols57} and atomic models for screening~\cite{Thomas_1927,Fermi_28,Moliere47}, modern studies concentrate on two tightly interwoven challenges: (i) the systematic inclusion of Coulomb distortion beyond the first Born approximation in a form that remains analytically tractable, and (ii) a faithful representation of atomic screening that smoothly covers charge states from neutral atoms to highly stripped ions. Both challenges are essential for constructing a single method that can reproduce experiments and benchmark calculations across the largest possible range of parameters.

The theoretical basis for such developments is the scattering-theory framework in which bremsstrahlung cross sections follow from Fermi’s golden rule~\cite{Fermi1950Nuclear,Schiff1968QM} (see Sec.~\ref{sec:Unscreened-Electron-ion-bremsstr}). In this paper we restrict attention to the thin-target, single-collision regime for electron energies up to a few tens of MeV, where quantum-coherence suppression effects (e.g. Landau–Pomeranchuk–Migdal~\cite{Landau_Pomeranchuk_1953a,Landau_Pomeranchuk_1953b,Migdal_1956}) are negligible. We note that fully rigorous S-matrix (Furry-picture) treatments~\cite{Furry_1951, Jauch_Rolich_1955}, coupled to double Dirac partial-wave (PW) expansions (the route pioneered by Tseng \& Pratt \cite{TsengPratt71}, and developed by Pratt and collaborators \cite{LeePratt76, Pratt_1977}) provide the most accurate nuclear-bremsstrahlung results available, but they require heavy numerical machinery and become prohibitive as the number of required PW increases exponentially with energy~\cite{pra75, YerSurzh, Jakubassa_brem}. We do not consider that PW S-matrix route in the present work. Instead, we focus on analytic and semi-analytic expressions that are directly compatible with our screening ansatz.

A recurring limitation of analytic bremsstrahlung treatments is the incorporation of screening. Classical prescriptions (Molière parameterizations~\cite{Moliere47}, Thomas–Fermi–like screening lengths~\cite{Thomas_1927,Fermi_28}, and tabulated atomic form factors~\cite{Hubbell,sal87}) describe neutral-atom behavior but do not by themselves provide a compact analytic route to interpolate across charge states. Alternative analytic screening laws have also been proposed, such as the Thomas–Fermi–Kirillov form (TFK), which replaces the Yukawa exponential by a power $n=3/2$~\cite{kir75}, see Eq.~\eqref{eq:atomic=000020form=000020factor}, but these remain difficult to integrate consistently into bremsstrahlung cross section formalisms. Model-potential approaches, most notably that of Avdonina and Pratt~\cite{avd93}, provide analytic screened atomic form factors (AFF) and explicitly address ionic charge states, demonstrating that screening significantly alters spectral and polarization observables. Their framework, however, relies on a specific one-parameter potential and is not easily generalized to reproduce high-quality tabulations or to interpolate smoothly between neutral and bare-ion limits. At the same time, investigations of Coulomb corrections emphasize that screening and higher-order distortion must be treated together to achieve percent-level accuracy in the application-relevant $\sim$1--100 MeV window and for moderate to high-$Z$ targets~\cite{roc72,Mangiarotti_2016}.

To overcome these difficulties, we introduce an analytic AFF representation built from a finite sum of Yukawa screening potentials~\cite{hau08, sav23}. Each term admits a simple analytic momentum-space component suitable for insertion into differential cross sections, while the linear combination can be fitted to high-quality atomic data (Dirac--Hartree--Fock--Slater (DHFS) fits~\cite{sal87,GAUSSIAN,GRASP} or Hubbell tabulations~\cite{Hubbell}) and adjusted to account for removed bound charge, thereby extending from neutral atoms to fully stripped ions.
Practically, screening is implemented by inserting the fitted multi-Yukawa (MY) form factor into the Bethe--Heitler (BH) differential cross section while higher-order Coulomb distortion is retained through the Roche–Ducos–Proriol (RDP) expression for the unscreened contribution~\cite{roc72}. Following the Olsen--Maximon--Wergeland (OMW) additivity rule \cite{ols57} (see Sec.~\ref{sec:Unscreened-Electron-ion-bremsstr}), the screened doubly differential cross section is obtained as the sum of the unscreened expression in a pure Coulomb field, and a screening correction in the first Born approximation, such that both neutral-atom and bare-nucleus limits are preserved. The result is a compact, rapidly evaluable expression for the doubly differential cross section (DDCS) that is inexpensive compared with full PW models.

The motivation for such a compact and accurate treatment spans several application domains. Relativistic electrons physics in nuclear fusion experiments~\cite{pey08b}, radiation-safety calculations, and astrophysical radiative-transfer problems~\cite{Kawakami_88, Chluba_2020}, among others, all require reliable bremsstrahlung cross sections for partially ionized targets. Recent advances in unified atomic physics models for Fokker–Planck solvers~\cite{sav23}, underline the need for analytic parameterizations that interpolate smoothly across charge states without reliance on large precomputed databases such as open-ADAS~\cite{sum04}. The present scheme is designed for direct integration into such kinetic and transport frameworks, offering fast evaluation and high accuracy.

The aims of the paper are therefore: (i) to present the multi-Yukawa AFF representation as a practical model for screening across ionization degrees; (ii) to derive bremsstrahlung differential cross sections obtained by combining this representation with a Sommerfeld--Maue-based formalism and consistent Coulomb corrections following the OMW additivity rule; and (iii) to validate the resulting scheme through systematic comparisons with existing models and experiments~\cite{Motz55,Starfelt56,Rester67,ResterEdmonson72,pra75}. The rest of the paper is organized as follows. Section~\ref{sec:Unscreened-Electron-ion-bremsstr} recalls the unscreened bremsstrahlung formalism from Fermi’s golden rule, and introduces the unscreened components of the model, namely the RDP and unscreened BH differential cross sections. Section~\ref{sec:Screened-Electron-ion-bremsstrah} defines the MY AFF and incorporates it in the derivation of the DDCS. Section~\ref{sec:Results} presents the numerical results and comparisons with experiments, and Section~\ref{sec:conclusions} summarizes conclusions and outlines future applications.

\section{Theoretical Framework for Electron–Ion Bremsstrahlung}\label{sec:Unscreened-Electron-ion-bremsstr}

In the present work, we assume that the reader is familiar with the theory of bremsstrahlung, and in particular with how the differential cross sections for bremsstrahlung interactions are derived. We recall that all formulae considered in this work originate from Fermi's \emph{golden rule} \cite{Fermi1950Nuclear,Schiff1968QM} that we remind here: 
\begin{align}
\frac{d^{3}\sigma_{e-i}}{dkd\Omega_{k}d\Omega_{p}} &:= \frac{\alpha}{(2\pi)^4}\lambdabar_C^2\frac{E_0Epk}{p_0}\frac{1}{2}\sum_{s,s_0,\mathbf{e^*}}|M|^2\label{eq:GoldenRule}\\
M&:=\int \psi^{\dagger}(\mathbf{r})(\boldsymbol{\alpha}\mathbf{\cdot e^*})e^{-i\bm{k\cdot r}}\psi_0(\mathbf{r})d^3\mathbf{r}.
\end{align}
The symbols involved are the usual for the context: $\sigma_{e-i}$ denotes the cross section, $\alpha$ is the fine structure constant and $\lambdabar_C=\hbar/m_ec$ the reduced Compton wavelength of the electron. $\Omega_k,\Omega_p$ denote the solid angles of emission for the photon and scattering of the electron, respectively, while $k$ is the radiated photon energy. The electron final and initial energies are $E, E_0$, respectively, and the same applies for the momenta $p,p_0$ and wave functions $\psi,\psi_0$. The squared matrix elements amplitudes $|M|^2$ are summed over the final and initial electron spin states $s,s_0$, as well as polarization vector of the photon $\mathbf{e}$. The reason for this averaging is that none of these quantities are observed. Finally, $\boldsymbol{\alpha}$ denotes the Dirac matrices. Note that throughout the work, energies and momenta are expressed in relativistic units, normalized to $m_ec^2$ and $m_ec$, respectively. Coordinates are expressed in units of the reduced Compton wavelength $\lambdabar_C$ such that wave factors take the form $\exp(i\boldsymbol{k}\cdot \boldsymbol{r})$. 

The triply-differential cross section (TDCS) in Eq.~\eqref{eq:GoldenRule} involves an integral over all space of the initial and final electron wave functions. In the vicinity of the nucleus, the nuclear (Coulomb) effects are important, while further from it, the electrons bound to the nucleus may screen the potential of the latter, so that a screening correction to the pure Coulomb field cross section ought to be considered. Olsen, Maximon and Wergeland noted that for the DDCS, i.e. after integrating out the electron's degrees of freedom, and for energies greater than the electron rest mass, the screening correction to the Coulomb cross section is additive \cite{ols57}: 
\begin{widetext}
\begin{equation}
\left(\frac{d^{2}\sigma_{e-i}^{Z_{s}}}{dkd\Omega_{k}}\right)^{\text{s}}\simeq\left(\frac{d^{2}\sigma_{e-i}^{Z}}{dkd\Omega_{k}}\right)_{\text{Coul}}^{\text{us}}
+\left[\left(\frac{d^{2}\sigma_{e-i}^{Z_{s}}}{dkd\Omega_{k}}\right)_{\text{Born}}^{\text{s}}-\left(\frac{d^{2}\sigma_{e-i}^{Z}}{dkd\Omega_{k}}\right)_{\text{Born}}^{\text{us}}\right].\label{eq:OMW}
\end{equation}
\end{widetext}
The latter is the so-called OMW \emph{additivity rule}. The term in brackets is the screening correction. The superscripts ``s'' and ``us'' mean screened and unscreened, respectively. An illustrating figure is presented at the beginning of Section~\ref{sec:Results}, while a discussion on the validity of the OMW rule can be found in~\cite{JakubassaMangia_2019}. We now introduce the expressions we will consider for the terms from Eq.~\eqref{eq:OMW}.

\subsection{Bethe--Heitler Cross Section\protect\label{subsec:Bethe-Heitler-cross-section}}

As discussed in the introduction to this paper, the triply differential cross section (differential in the photon energy, the photon emission solid angle, and the electron scattering solid angle) as derived by Bethe and Heitler~\cite{bet34} and labeled 1BS in Ref.~\cite{koc59}, provides
the general expression for bremsstrahlung in the first
Born approximation. It is given by
\begin{align}
\left(\frac{d^{3}\sigma_{e-i}^{Z_{s}}}{d{k}d\Omega_{k}d\Omega_{p}}\right)_{\text{BH}}=&\alpha\left(\frac{r_{e}}{2\pi}\right)^{2}Z^2\left(1-F_{Z_{s}}\left({q}\right)\right)^{2}\\
&\times\frac{{p}}{{k}p_{0}{q}^{4}}A^{\text{BH}},\label{eq:cross-section=0000201BS}
\end{align}
with $r_e$ the classical electron radius, and where $F_{Z_{s}}$ is the atomic form factor (AFF). The latter is set to zero in the unscreened case, and we retain this expression for the unscreened bremsstrahlung term, i.e. the third on the right-hand side of Eq.~\eqref{eq:OMW}. The recoil momentum transferred to the nucleus, ${q}$, is given by the relation
\begin{align}
\boldsymbol{q}=\boldsymbol{p}_0-\boldsymbol{p}-\boldsymbol{k},\label{eq:recoil}
\end{align}
such that
\begin{align}
{q}^{2}=&{p}_{0}^{2}+{p}^{2}+{k}^{2}-2{p}_{0}{k}\cos\theta_{0}+2{p}{k}\cos\theta\\
&-2{p}{p}_{0}\left(\cos\theta\cos\theta_{0}+\sin\theta\sin\theta_{0}\cos\phi\right),\label{eq:recoil=000020momentum}
\end{align}
where $\theta_{0}$ and $\theta$ denote the angles between the emitted photon and the incoming and scattered electron momenta, respectively, and $\phi$ is the angle between the planes $(\boldsymbol{p}_0,\boldsymbol{k})$ and $(\boldsymbol{p},\boldsymbol{k})$. Here, and from now on, $Z_{s}:=Z - N_{s}$ represents the effective ionic charge of the species, that is, the atomic number minus the number of bound electrons. The term $A^{\text{BH}}$
in Eq.~\eqref{eq:cross-section=0000201BS} is given by
\begin{equation}
A^{\text{BH}}=\sum_{n=1}^{4}A_{n}^{\text{BH}},\label{eq:An=000020series}
\end{equation}
where 
\begingroup
\mathtoolsset{showonlyrefs=false}
\begin{subequations}\label{eq:ABH_all}
\begin{align}
A_{1}^{\text{BH}} & =  \frac{{p}^{2}\sin^{2}\theta}{\left({E}-{p}\cos\theta\right)^{2}}\left(4{E}_{0}^{2}-{q}^{2}\right)\label{eq:ABH_1}\\
A_{2}^{\text{BH}} & =  \frac{{p}_{0}^{2}\sin^{2}\theta_{0}}{\left({E}_{0}-{p}_{0}\cos\theta_{0}\right)^{2}}\left(4{E}^{2}-{q}^{2}\right)\label{eq:ABH_2}\\
A_{3}^{\text{BH}} & =  2{k}^{2}\frac{{p}^{2}\sin^{2}\theta+{p}_{0}^{2}\sin^{2}\theta_{0}}{\left({E}-{p}\cos\theta\right)\left({E}_{0}-{p}_{0}\cos\theta_{0}\right)}\label{eq:ABH_3}\\
A_{4}^{\text{BH}} & =  -\frac{2{p}{p}_{0}\sin\theta\sin\theta_{0}\cos\phi}{\left({E}-{p}\cos\theta\right)\left({E}_{0}-{p}_{0}\cos\theta_{0}\right)}\left(2{E}^{2}+2{E}_{0}^{2}-{q}^{2}\right).\label{eq:ABH_4}
\end{align}
\end{subequations}
\endgroup

For certain combinations of energies, momenta and angles, the expressions~\eqref{eq:ABH_1}-\eqref{eq:ABH_4} exhibit near exact cancellations, and careful numerical treatment is required when evaluating the expression as is. This issue has been discussed in detail by Maximon \textit{et al}. \cite{MAXIMON1987189}, who proposed a refactored formulation to overcome this caveat. In the present work, this difficulty does not arise in that the integrations performed for both the unscreened and screened cases do not exhibit such canceling terms.

In the unscreened limit ($F_{Z_{s}}=0$), Eq.~(\ref{eq:cross-section=0000201BS}) can be integrated analytically over all the possible scattering directions, recalling that $d\Omega_{p}=\sin\theta d\theta d\phi$. The integration has first been done by Sauter~\cite{Sauter34}, and the result is denoted 2BN in Ref.~\cite{koc59}. We will refer to the DDCS without screening as the Sauter (S) formula, even though it comes from integration of the Bethe--Heiler TDCS. It reads  
\begin{align}
\left(\frac{d^{2}\sigma_{e-i}^{Z}}{d{k}d\Omega_{k}}\right)_{\text{S}}^{\text{us}}&=\int \left(\frac{d^{3}\sigma_{e-i}^{Z}}{d{k}d\Omega_{k}d\Omega_{p}}\right)_{\text{BH}}\!d\Omega_{p}\\
&=\alpha r_{e}^{2}\frac{Z^{2}}{8\pi}\frac{{p}}{{k}{p}_{0}}\sum_{n=1}^{10}B_{n}^{\text{S}},\label{eq:cross-section=0000202BN-1}
\end{align}
where 
\begin{align}
B_{1}^{\text{S}} & =  \frac{8\sin^{2}\theta_{0}}{{p}_{0}^{2}\Delta_{0}^{4}}\left(2{E}_{0}^{2}+1\right)\nonumber\\
B_{2}^{\text{S}} & =  -\frac{2\left(5{E}_{0}^{2}+2{E}{E}_{0}+3\right)}{{p}_{0}^{2}\Delta_{0}^{2}}\nonumber\\
B_{3}^{\text{S}} & =  -\frac{2\left({p}_{0}^{2}-{k}^{2}\right)}{Q^{2}\Delta_{0}^{2}}\nonumber\\
B_{4}^{\text{S}} & =  \frac{4{E}}{{p}_{0}^{2}\Delta_{0}}\nonumber\\
B_{5}^{\text{S}} & =  \frac{L}{{p}{p}_{0}}\frac{4{E}_{0}\sin^{2}\theta_{0}\left(3{k}-{p}_{0}^{2}{E}\right)}{{p}_{0}^{2}\Delta_{0}^{4}}\nonumber\\
B_{6}^{\text{S}} & =  \frac{L}{{p}{p}_{0}}\frac{4{E}_{0}^{2}\left({E}_{0}^{2}+{E}^{2}\right)}{{p}_{0}^{2}\Delta_{0}^{2}}\nonumber\\
B_{7}^{\text{S}} & =  \frac{L}{{p}{p}_{0}}\frac{2-2\left(7{E}_{0}^{2}-3{E}{E}_{0}+{E}^{2}\right)}{{p}_{0}^{2}\Delta_{0}^{2}}\nonumber\\
B_{8}^{\text{S}} & =  \frac{L}{{p}{p}_{0}}\frac{2{k}\left({E}_{0}^{2}+{E}{E}_{0}-1\right)}{{p}_{0}^{2}\Delta_{0}}\nonumber\\
B_{9}^{\text{S}} & =  -\frac{4\epsilon}{{p}\Delta_{0}}\nonumber\\
B_{10}^{\text{S}} & =  \frac{\epsilon^{Q}}{{p}Q}\left[\frac{4}{\Delta_{0}^{2}}-\frac{6{k}}{\Delta_{0}}-\frac{2{k}\left({p}_{0}^{2}-{k}^{2}\right)}{Q^{2}\Delta_{0}}\right]\nonumber,
\end{align}
with 
\begin{align}
L=\ln\left(\frac{{E}{E}_{0}-1+{p}{p}_{0}}{{E}{E}_{0}-1-{p}{p}_{0}}\right)\nonumber,
\end{align}
$\Delta_{0}={E}_{0}-{p}_{0}\cos\theta_{0}$, and $Q^{2}={p}_{0}^{2}+{k}^{2}-2{p}_{0}{k}\cos\theta_{0}$.
In addition, 
\begin{align}
\epsilon=\ln\left(\frac{{E}+{p}}{{E}-{p}}\right)\quad \text{and}\quad \epsilon^{Q}=\ln\left(\frac{Q+{p}}{Q-{p}}\right).
\end{align}

The first Born approximation, and thus the Bethe--Heitler--Sauter cross section of Eq.~(\ref{eq:cross-section=0000202BN-1}), is valid only when $\alpha Z\!\ll\!(\beta_{0},\beta)$, a very restrictive condition.
Under this approximation, the Born formulas yield cross sections larger than experimental values in the soft-photon limit ($k\!\ll\!E$), but fall below them in the tip (high frequency) region, where the predicted cross section goes to zero as the photon energy reaches the incident energy.

\subsection{Sauter--Elwert Cross Section}\label{subsec:Bethe-Heitler-Elwert-cross-secti}
As discussed previously, in the short-wavelength limit ($k \!\uparrow\! E$), 
the Born-approximation doubly differential cross section (DDCS) fails, 
vanishing as the photon energy approaches the incident electron energy. 
However, the cross section remains finite if it is multiplied by an appropriate 
Coulomb correction factor derived by Elwert and Haug~\cite{elw69},
\begin{align}\label{eq:Elwertfactor}
F_{\text{E}}\!\left({p}_{0},{p},Z\right)
  =\frac{\xi}{\xi_{0}}
   \frac{1-\exp\!\left(-2\pi\xi_{0}\right)}{1-\exp\!\left(-2\pi\xi\right)},
\end{align}
where $\xi_{0}=\alpha Z{E}_{0}/{p}_{0}=\alpha Z\beta_{0}^{-1}$ and
$\xi=\alpha Z\beta^{-1}$.
The resulting Sauter--Elwert (SE) cross section is then defined as
\begin{align}\label{eq:cross-sectionBHE}
\left(\frac{d^{2}\sigma_{e-i}^{Z}}{d{k}\,d\Omega_{k}}\right)_{\!\text{SE}}^{\text{us}}
  :=F_{\text{E}}\!\left({p}_{0},{p},Z\right)
   \left(\frac{d^{2}\sigma_{e-i}^{Z}}{d{k}\,d\Omega_{k}}\right)_{\!\text{S}}^{\text{us}} .
\end{align}

Physically, the Elwert factor $F_{\text{E}}$ accounts for the Coulomb distortion 
of the initial and final electron wave functions. 
It acts at the triply differential level where it modifies the local 
transition probability density. When $F_{\text{E}}$ is applied to the TDCS and one integrates over the outgoing 
electron angle $\Omega_{p}$ to obtain the DDCS, the resulting expression becomes 
essentially independent of the photon direction $\Omega_{k}$ in the 
neighborhood of the long-wavelength limit. As $\xi_{0}$ decreases, this angular 
independence extends progressively toward the short-wavelength limit~\cite{ElwertDissert}.
This near-constancy allows $F_{\text{E}}$ to be treated as a global multiplicative 
factor that can be pulled out of the angular integration, yielding the compact form 
in Eq.~\eqref{eq:cross-sectionBHE}. 
At shorter wavelengths, however, where the outgoing electron becomes slow and 
Coulomb distortion is strongest, this approximation breaks down and 
$F_{\text{E}}$ must in principle be retained inside the angular integrals. 
Elwert and Haug~\cite{elw69} addressed this by computing numerical 
Coulomb factors that extend the validity of the correction into the 
short-wavelength regime.

The Sauter--Elwert cross section is valid in the non-relativistic regime 
($\xi_{0}\!\ll\!1$) and for sufficiently small atomic numbers $(Z\lesssim 26)$. Unlike the Sauter 
result, it does not vanish at the tip region. For a more detailed discussion on Coulomb factors, we refer to~\cite{ElwertDissert, elw69}.

For highly relativistic incident electrons, and higher atomic numbers, the Sauter--Elwert corrected cross section becomes inadequate and more advanced numerical treatments of the Coulomb field, such 
as the Roche--Ducos--Proriol cross section, are required.

\subsection{Roche--Ducos--Proriol Cross Section\protect\label{subsec:Roche-Ducos-Proriol-cross-sectio}}
For high-$Z$ species ($Z>26$), the Sauter expression corrected within the first Born approximation (with Elwert factors) is no longer reliable, particularly in the tip region (see Sec.~\ref{subsec:Bethe-Heitler-Elwert-cross-secti}). In such cases one must turn to formulations that treat the Coulomb field more precisely. The first analytic progress in this direction was made by Elwert and Haug, who derived their cross section using Furry--Sommerfeld--Maue (FSM) (usually simply called Sommerfeld--Maue (SM)) wave functions~\cite{Furry34_SM,SommerfeldMaue_35} for the electron in the pure Coulomb potential~\cite{elw69}. This approach was later taken to the next order by Roche, Ducos, and Proriol, who added the dominant higher-order terms beyond Elwert--Haug~\cite{roc72}. Despite its extensive use in the literature~\cite{hau08, Mangiarotti_2016, MangiarottiMartins_2017, MangiarottiJakubassa_17}, the RDP cross section as introduced in the original source~\cite{roc72} is not consistent, as its mixed-order character (combining third- and fourth-order contributions) does not necessarily lead to improved accuracy, as originally claimed. Although the mixed-order aspect of the initial formalism had been justified previously~\cite{hau08, Mangiarotti_2016, MangiarottiMartins_2017, MangiarottiJakubassa_17}, the resulting inconsistency was later identified and corrected at the level of the radiation matrix element by Jakubassa-Amundsen and Mangiarotti~\cite{JakubassaMangia_2019}. Here, we extend this correction to the level of the triply-differential cross section, yielding a consistently truncated third-order expression. The corrected RDP formalism provides the most accurate closed-form description of bremsstrahlung in a pure Coulomb field in the energy range from a few to several tens of MeV, and the integrated RDP TDCS is the expression we employ for the first term on the right-hand side of Eq.~\eqref{eq:OMW}. The RDP cross section reads
\begin{align}
\left(\frac{d^{3}\sigma_{e-i}^{Z}}{d{k}d\Omega_{k}d\Omega_{p}}\right)_{\text{RDP}}^{\text{us}}:=\frac{d^{3}\sigma_{\text{RDP},1}^{Z}}{d{k}d\Omega_{k}d\Omega_{p}}+\frac{d^{3}\sigma_{\text{RDP},2}^{Z}}{d{k}d\Omega_{k}d\Omega_{p}}\label{eq:RDP_tot}.
\end{align}
The first term on the right-hand side is identical to the cross section derived by Elwert
and Haug \cite{elw69}, 
\begin{align}
\frac{d^{3}\sigma_{\text{RDP},1}^{Z}}{d{k}d\Omega_{k}d\Omega_{p}}\equiv\frac{d^{3}\sigma_{\text{EH}}^{Z}}{d{k}d\Omega_{k}d\Omega_{p}},\label{eq:EH=000020cross-section}
\end{align}
while the second term, using the Bethe-Maximon method \cite{bet54},
describes higher-order corrections. From Refs. \cite{roc72,hau08},
\begin{align}
\frac{d^{3}\sigma_{\text{RDP},1}^{Z}}{d{k}d\Omega_{k}d\Omega_{p}} & =  \frac{\alpha Z^{2}}{\pi^{2}}r_{e}^{2}\frac{{k}{p}}{{p}_{0}}N\left\{ \left[{E}_{0}{E}-1\right.\right.\nonumber \\
 & -  \left.\left(\hat{\bm{k}}\cdot{\bm{p}}_0\right)\left(\hat{\bm{k}}\cdot{\bm{p}}\right)\right]\left|J_{1}\right|^{2}\nonumber \\
 & + \left[{E}_{0}{E}+1+\left(\hat{\bm{k}}\cdot{\bm{p}}_0\right)\left(\hat{\bm{k}}\cdot{\bm{p}}\right)\right]\left(\left|\bm{J}_{2}\right|^{2}+\left|\bm{J}_{3}\right|^{2}\right)\nonumber \\
 & +  2\Re\left[\left(\bm{J}_{3}^{*}-\bm{J}_{2}^{*}\right)\cdot\left\{ {\bm{p}}_0\left(\bm{J}_{2}\cdot\hat{\bm{k}}\right)\left({\bm{p}}\cdot\hat{\bm{k}}\right)\right.\right.\nonumber \\
 & -  \left.{\bm{p}}\left(\bm{J}_{3}\cdot\hat{\bm{k}}\right)\left({\bm{p}}_0\cdot\hat{\bm{k}}\right)\right\} \nonumber \\
 & -  \left({E}_{0}{E}+1+{\bm{p}}_0\cdot{\bm{p}}\right)\left(\bm{J}_{2}\cdot\hat{\bm{k}}\right)\left(\bm{J}_{3}^{*}\cdot\hat{\bm{k}}\right)\nonumber \\
 & +  \left(\bm{J}_{2}\cdot{\bm{p}}_0\right)\left(\bm{J}_{3}^{*}\cdot{\bm{p}}\right)-\left(\bm{J}_{2}\cdot{\bm{p}}\right)\left(\bm{J}_{3}^{*}\cdot{\bm{p}}_0\right)\nonumber \\
 & +  {E}_{0}J_{1}^{*}\left\{ \bm{J}_{3}\cdot{\bm{p}}-\left(\bm{J}_{2}\cdot\hat{\bm{k}}\right)\left({\bm{p}}\cdot\hat{\bm{k}}\right)\right\} \nonumber \\
 & +  {E}J_{1}^{*}\left\{ \bm{J}_{2}\cdot{\bm{p}}_0-\left(\bm{J}_{3}\cdot\hat{\bm{k}}\right)\left({\bm{p}}_0\cdot\hat{\bm{k}}\right)\right\} \nonumber \\
 & +  \left.\left.\left(\bm{J}_{2}\cdot\bm{J}_{3}^{*}\right)\left\{ {\bm{p}}_0\cdot{\bm{p}}-\left(\hat{\bm{k}}\cdot{\bm{p}}_0\right)\left(\hat{\bm{k}}\cdot{\bm{p}}\right)\right\} \right]\right\}, \label{eq:sigma_ei_rdp_1}
\end{align}
while the consistent third-order RDP term reads
\begin{align}
\frac{d^{3}\sigma_{\text{RDP},2}^{Z}}{d{k}d\Omega_{k}d\Omega_{p}}  &=  \alpha^{2}Z^{3}r_{e}^{2}\frac{{k}{p}\left({\bm{q}}\cdot{\bm{k}}\right)}{{p}_{0}{q}D_{0}D}N\\
&\times\left\{ \frac{2}{\pi}\Re\left[\left(\left\{ {E}_{0}{E}-1-\left(\hat{\bm{k}}\cdot{\bm{p}}_0\right)\left(\hat{\bm{k}}\cdot{\bm{p}}\right)\right\} J_{1}\right.\right.\right.\nonumber \\
 & +  {E}_{0}\left\{ \bm{J}_{3}\cdot{\bm{p}}-\left(\bm{J}_{2}\cdot\hat{\bm{k}}\right)\left({\bm{p}}\cdot\hat{\bm{k}}\right)\right\} \nonumber \\
 & +  \left.\left.{E}\left\{ \bm{J}_{2}\cdot{\bm{p}}_{0}-\left(\bm{J}_{3}\cdot\hat{\bm{k}}\right)\left({\bm{p}}_{\mathrm{0}}\cdot\hat{\bm{k}}\right)\right\} \right)e^{i\Phi}\right]. \label{eq:sigma_ei_rdp_2}
\end{align}
In Eqs.~\eqref{eq:sigma_ei_rdp_1} and~\eqref{eq:sigma_ei_rdp_2}, ${\bm{k}}={k}\bm{\hat{k}}$ and 
\begin{align}
N=\frac{4\pi^{2}a_{0}a}{\left(e^{2\pi a_{0}}-1\right)\left(e^{2\pi a}-1\right)},
\end{align}
where $a_{0}=\left({E}_{0}/{p}_{0}\right)\alpha Z_{s}$ and
$a=\left({E}/{p}\right)\alpha Z_{s}$, while 
\begin{align}
D_{0}  &=  2\left({E}_{0}{k}-{\bm{p}}_{0}\cdot{\bm{k}}\right) \quad  D  =  2\left({E}{k}-{\bm{p}}\cdot{\bm{k}}\right)\\
  \mu&=2\left({E}_{0}{E}+{p}_{0}{p}-1\right) \quad \text{and} \quad x=1-\frac{D_{0}D}{\mu{q}^{2}}.
\end{align}
The phase $\Phi$ in Eq.~\eqref{eq:sigma_ei_rdp_2} is given
by 
\begin{align}
\Phi=a_{0}\ln\left(\frac{{q}^{2}}{D}\right)-a\ln\left(\frac{\mu}{D}\right),
\end{align}
and the coefficients $J_{1}$, $\bm{J}_{2}$ and $\bm{J}_{3}$ are 
\begin{align}
J_{1}  =&  2\left(\frac{{E}}{D_{0}}-\frac{{E}_{0}}{D}\right)\frac{V+iaxW}{{q}^{2}}\nonumber\\
&+2i\frac{\left(1-x\right)W}{D_{0}D}\left[{E}_{0}a\left(\frac{\mu}{D}-1\right)-{E}a_{0}\left(\frac{\mu}{D_{0}}+1\right)\right]\nonumber\\
\bm{J}_{2}  =&  \frac{V+iaxW}{D{q}^{2}}{\bm{q}}-\frac{ia\left(1-x\right)W}{D_{0}D}\left[\left(\frac{\mu}{D}-1\right){\bm{q}}-\frac{{\bm{P}}}{{p}_{0}}\right]\nonumber\\
\bm{J}_{3}  =&  \frac{V+iaxW}{D{q}^{2}}{\bm{q}}-\frac{ia_{0}\left(1-x\right)W}{D_{0}D}\left[\left(\frac{\mu}{D_{0}}+1\right){\bm{q}}-\frac{{\bm{P}}}{{p}}\right].\nonumber
\end{align}
As for $V={}_{2}F_{1}\left(-ia_{0},ia;1;x\right)$ and $W={}_{2}F_{1}\left(1-ia_{0},1+ia;2;x\right)$, $_{2}F_{1}$ denotes the Gaussian hypergeometric function with
argument $x\in [ 0,1 ]$ \cite{abr70}, while 
\begin{align}
{\bm{P}}={p}_{0}{\bm{p}}+{p}{\bm{p}}_{\mathrm{0}}.
\end{align}

Due to their high complexity, the expressions~\eqref{eq:sigma_ei_rdp_1} and~\eqref{eq:sigma_ei_rdp_2} can only be integrated over $\Omega_p$ numerically. The evaluation of two hypergeometric functions at each point renders the computation demanding; for computational efficiency, the latter are implemented in the form of series expansions, and the results have been benchmarked against an arbitrary-precision library~\cite{arb_flint}. Details on the numerical implementation can be found in the source repository~\cite{brem_cross_sec_release}.

The use of FSM wave functions is valid as long as 
\begin{align}
\frac{(\alpha Z)^2}{r}\sin\left(\frac{\vartheta_{pr}}{2}\right)\ll 1,\label{eq:FSM_cond}
\end{align}
where $\vartheta_{pr}$ is the angle between ${\bm{p}}$ and $\mathbf{r}$, and $r=\|\mathbf{r}\|$, the electron--nucleus separation~\cite{haug2004elementary}.  This condition is satisfied for low atomic numbers at all angles and energies, but its validity is restricted to small angles for higher $Z$. In practice this approximation is sufficient since photon emission is concentrated in the forward direction at higher electron energies. The implications of this condition are further discussed in Sec.~\ref{sec:Results}.

\section{Screened Electron–Ion Bremsstrahlung: The Multi-Yukawa Model}\label{sec:Screened-Electron-ion-bremsstrah}
To evaluate the screening correction term in Eq.~\eqref{eq:OMW}, the screened DDCS is derived in the first Born approximation. In this section, the latter is constructed by introducing an atomic form factor based on the multi-Yukawa model and inserting it into the Bethe--Heitler TDCS. Integration over the electron scattering solid angle is then carried out analytically to yield the corresponding DDCS.

\subsection{Atomic Form Factor\protect\label{subsec:Atomic-form-factor}}
From the Thomas--Fermi--Kirillov (TFK) or Yukawa (Y) approximate atomic
potential models~\cite{kir75,avd93} the form
factor for the ion $Z_{s}$, $F_{Z_{s}}\left({q}\right)$, may be expressed in the
simple form\footnote{Note that the present definition of the form factor differs from Refs.~\cite{kir75,sav23} by an overall factor of $1/Z$. Here we adopt the convention commonly used in the bremsstrahlung literature.}
\begin{equation}
F_{Z_{s}}\left({q}\right)=\frac{1}{Z}\frac{N_{s}}{1+\left({q}{a}_{Z_{s}}/2\right)^{n}},\label{eq:atomic=000020form=000020factor}
\end{equation}
where $n=3/2$ and $n=2$ for the TFK and Y potentials, respectively.
Here ${a}_{Z_{s}}=2\lambda_{Z_{s}}/\alpha$ is the effective screening length, with $\lambda_{Z_{s}}$ the screening parameter of the underlying Yukawa or TFK potential. It depends on the atomic model and the ionization state. In the small-${q}$ limit, the form factor reduces to $F_{Z_{s}}({q}) \simeq N_{s}/Z$, in agreement with charge conservation.

For an atomic potential represented as a sum of multiple Yukawa (MY) terms, the AFF~\eqref{eq:atomic=000020form=000020factor} takes the form
\beq
F_{Z_{s}}\left({q}\right)  =  \frac{N_{s}}{Z}\sum_{i}\frac{{A}_{s,i}}{1+\left({q}{a}_{Z_{s,i}}/2\right)^{2}},\label{eq:form=000020factor=000020moliere=0000203=000020exp-1}
\eeq
where $\{A_{s,i}\}_{i}$ denote the weights corresponding to each exponential for the different ionization levels. For normalization, we impose $\sum_{i}{A}_{s,i}=1$. This formulation provides an accurate description of screening effects, regardless of the ionization state or the atomic species. We refer to the work by Savoye-Peysson \textit{et al.} for a detailed description of the MY model and its underlying assumptions~\cite{sav23}. Integration over $\phi$ and $\theta$ of the outgoing electron 
yields the DDCS in photon energy and emission angle
\begin{equation}
\frac{d^{2}\sigma^{\text{MY}}_{Z_{s}}}{d{k}d\Omega_{k}}=\int_{0}^{\pi}\int_{0}^{2\pi}\frac{d^{3}\sigma^{\text{MY}}_{Z_{s}}}{d{k}d\Omega_{k} d\Omega_{p}} d\phi\sin\theta d\theta.\label{eq:cross-section=0000202BN}
\end{equation}

\subsection{Screened Cross Section\protect\label{subsec:Screened-cross-section}}
For a standard Yukawa ($n=2$) model, fully analytic expressions for the DDCS have been obtained~\cite{Fronsdal58,hau08}. In contrast, for the approximate Thomas--Fermi--Kirillov atomic model ($n=3/2$), the rational exponent prevents a closed-form expression, and a numerical integration is therefore required. 

In this section, a fully analytic expression is derived for a multi-Yukawa (MY) model using the MY series, 
$$\{{A}_{s,i}, a_{Z_{s,i}}\}_i,$$ 
determined via the generalized Salvat moment method for arbitrary ionization states~\cite{sal87,sav23}. 

Combining Eq.~\eqref{eq:cross-section=0000201BS} with the AFF Eq.~\eqref{eq:atomic=000020form=000020factor},
\begin{align}
&\frac{d^{3}\sigma_{Z_{s}}^{\text{MY}}}{d{k}d\Omega_{k}d\Omega_{p}}=   \alpha\left(\frac{r_{e}}{2\pi}\right)^{2}Z^{2} \times\label{eq:Dsigmabrem1}\\
 &  \left(1-\left(1-q_{Z_{s}}\right)\sum_{i}\frac{{A}_{s,i}}{1+\left({q}{a}_{Z_{s,i}}/2\right)^{2}}\right)^{2}\frac{{p}}{{k}{p}_{0}{q}^{4}}A^{\text{BH}},
\end{align}
where the term $q_{Z_{s}}:= Z_{s}/Z$, ranging between
$0$ and $1$, characterizes the level of ionization, $0$ being the neutral atom. Defining $b_{s,i}:=2/{a}_{Z_{s,i}}$, Eq.~\eqref{eq:Dsigmabrem1} may then be rewritten in the form
\begin{equation}
\frac{d^{3}\sigma_{Z_{s}}^{\text{MY}}}{d{k}d\Omega_{k}d\Omega_{p}}=\left(\sum_{i}{A}_{s,i}\frac{{q}^{2}+q_{Z_{s}}{b}_{s,i}^{2}}{{q}^{2}+{b}_{s,i}^{2}}\right)^{2}\frac{d^{3}\sigma_{Z}^{\text{BH}}}{d{k}d\Omega_{k}d\Omega_{p}}.\label{eq:dsigmabrem3}
\end{equation}
From now on, we omit the BH superscript on the Bethe--Heitler TDCS for the sake of readability. For a neutral atom, $q_{Z_{0}}=0$, so 
\begin{align}
\frac{d^{3}\sigma_{Z_{0}}^{\text{MY}}}{d{k}d\Omega_{k}d\Omega_{p}} & =  {q}^{4}\left(\sum_{i}\frac{{A}_{0,i}}{{b}_{0,i}^{2}+{q}^{2}}\right)^{2}\frac{d^{3}\sigma_{Z}}{d{k}d\Omega_{k}d\Omega_{p}},\label{eq:dsigmabrem3-neutral=000020atom}
\end{align}
a form already found in Ref.~\cite{hau08}. 

For partially ionized atoms, the complexity is considerably higher
because of the presence of the terms $\propto q_{Z_{s}}{A}_{s,i}{b}_{s,i}^{2}$. Let us define 
\begin{align}
{c}_{s,i}^{2}:= q_{Z_{s}}{b}_{s,i}^{2},\label{eq:ci}
\end{align}
which mixes ionization and screening, so the more symmetric form
\begin{align}
\frac{d^{3}\sigma_{Z_{s}}^{\text{MY}}}{d{k}d\Omega_{k}d\Omega_{p}}=\left(\sum_{i}{A}_{s,i}\frac{{c}_{s,i}^{2}+{q}^{2}}{{b}_{s,i}^{2}+{q}^{2}}\right)^{2}\frac{d^{3}\sigma_{Z}}{d{k}d\Omega_{k}d\Omega_{p}}\label{eq:dsigmabrem3-ci}
\end{align}
is obtained leveraging the positivity of $q_{Z_{s}}$. For the usual Y potential with a single exponential,
$i=1$ and since ${A}_{s,1}=1$ by definition,
\begin{align}
\frac{d^{3}\sigma_{Z_{s}}^{\text{Y}}}{d{k}d\Omega_{k}d\Omega_{p}} 
  &= \frac{d^{3}\sigma_{Z}}{d{k}d\Omega_{k}d\Omega_{p}}\{\}, \label{eq:screened-Y-cross-section-1} \\
\{\} &= {c}_{s}^{4} I_{2,0}\!\left({b}_{s},{q}\right) \\
     &+ 2{c}_{s}^{2} I_{2,1}\!\left({b}_{s},{q}\right)+ I_{2,2}\!\left({b}_{s},{q}\right),
     \label{eq:screened-Y-cross-section-2}
\end{align}
where 
\begin{equation}
I_{l,m}\left({b},{q}\right)=\frac{{q}^{2m}}{\left({b}^{2}+{q}^{2}\right)^{l}}.\label{eq:I2m}
\end{equation}
When screening effects are neglected, $b_{s}=c_{s}=0$, so only the last term in expression~\eqref{eq:screened-Y-cross-section-2} remains. In this case, the Bethe--Heitler TDCS is correctly recovered since $I_{2,2}\left(0,q\right)=1$.
Conversely, for neutral atoms with screening ($c_{0,s}=0$, $b_{0,s}\neq 0$) and
\begin{equation}
\frac{d^{3}\sigma_{Z_{0}}^{\text{Y}}}{d{k}d\Omega_{k}d\Omega_{p}}=I_{2,2}\left({b}_{0},{q}\right)\frac{d^{3}\sigma_{Z}}{d{k}d\Omega_{k}d\Omega_{p}},
\end{equation}
which is similar to the expression found in \cite{hau08}. 

Having established the robustness of the construction through multiple checks, we now derive the doubly-differential cross section for the general MY case. When several exponentials are considered, the screening factor 
\begin{align}
\left(\sum_{i}{A}_{s,i}\frac{{c}_{s,i}^{2}+{q}^{2}}{{b}_{s,i}^{2}+{q}^{2}}\right)^{2} \label{eq:screening_AFF_MY}
\end{align}
is fully expanded by developing the four terms of the product. The TDCS for a MY potential then reads
\begin{widetext}
\begin{align}
\frac{d^{3}\sigma_{Z_{s}}^{\text{MY}}}{d{k}\,d\Omega_{k}\,d\Omega_{p}} 
&= \frac{d^{3}\sigma_{Z}}{d{k}\,d\Omega_{k}\,d\Omega_{p}}  \\
&\times \biggl\{
\sum_{i} {A}_{s,i}^{2} \bigl[
    {c}_{s,i}^{4} I_{2,0}\left({b}_{s,i},{q}\right)
    + 2{c}_{s,i}^{2} I_{2,1}\left({b}_{s,i},{q}\right)
    + I_{2,2}\left({b}_{s,i},{q}\right)
\bigr]  \\
&\quad + \sum_{i \ne i'} \frac{{A}_{s,i}\,{A}_{s,i'}}{%
    \left({b}_{s,i'}^{2}-{b}_{s,i}^{2}\right)}
\bigl[
    {c}_{s,i}^{2} {c}_{s,i'}^{2}
    \bigl(I_{1,0}\left({b}_{s,i},{q}\right)
        - I_{1,0}\left({b}_{s,i'},{q}\right)\bigr)  \\
&\quad\quad + \left({c}_{s,i}^{2} + {c}_{s,i'}^{2}\right)
    \bigl(I_{1,1}\left({b}_{s,i},{q}\right)
        - I_{1,1}\left({b}_{s,i'},{q}\right)\bigr)  \\
&\quad\quad + I_{1,2}\left({b}_{s,i},{q}\right)
           - I_{1,2}\left({b}_{s,i'},{q}\right)
\bigr]
\biggr\}.
\label{eq=screened_MY_cross_section}
\end{align}
\end{widetext}
The latter expression~\eqref{eq=screened_MY_cross_section} must be integrated over the scattering directions $\Omega_p$ to obtain the screened, ionized doubly differential cross section. It is easily verified that upon setting $c_{s,i}=0$, the neutral case from \cite{hau08} is recovered and that in the unscreened limit ${b}_{s,i}^{2}\downarrow 0$, the Bethe--Heitler TDCS is retrieved.
Let us define the double integrals $\mathcal{H}_{l,m}$
for $m\in\left\{ 0,1,2\right\} $ and $l\in\left\{ 1,2\right\}$,
\allowdisplaybreaks{
\begin{align}
\mathcal{H}_{l,m}\left({b}, {q}\right) & :=  \int_{0}^{\pi}\int_{0}^{2\pi}\frac{d^{3}\sigma_{Z}}{d{k}d\Omega_{k}d\Omega_{p}}I_{l,m}\left({b},{q}\right) d\phi \sin \theta  d\theta,\nonumber \\
 & =  \int_{0}^{\pi}\int_{0}^{2\pi}\frac{d^{3}\sigma_{Z}}{d{k}d\Omega_{k}d\Omega_{p}}\frac{{q}^{2m}}{({b}^{2}+{q}^{2})^{l}}
d\phi \sin\theta d\theta. \label{eq:Hm}
\end{align}
} 
Note that in order to determine the screened DDCS, only the integrals $\mathcal{H}_{1,m}\left({b}\right)$ ought to be determined explicitly, since $\mathcal{H}_{2,m}\left({b}\right)=-\partial\mathcal{H}_{1,m}\left({b}\right)/\partial {b}^{2}$,
as a consequence of the relation
\allowdisplaybreaks{
\begin{align}
\frac{\partial I_{1,m}({b})}{\partial{b}^{2}}=-\frac{{q}^{2m}}{({b}^{2}+{q}^{2})^{2}}=-I_{2,m}({b}).\label{eq:H2m=000020vs=000020H1m}
\end{align}}
Thus the DDCS in presence of screening, with the MY model, becomes
\begin{widetext}
\begin{align}
\frac{d^{2}\sigma_{Z_{s}}^{\text{MY}}}{d{k}d\Omega_{k}} & =  \sum_{i}{A}_{s,i}^{2}\left[{c}_{s,i}^{4}\mathcal{H}_{2,0}\left({b}_{s,i}\right)+2{c}_{s,i}^{2}\mathcal{H}_{2,1}\left({b}_{s,i}\right)+\mathcal{H}_{2,2}\left({b}_{s,i}\right)\right] \nonumber\\
 & +   \sum_{i\neq i'}\frac{{A}_{s,i}{A}_{s,i'}}{\left({b}_{s,i'}^{2}-{b}_{s,i}^{2}\right)}\left[{c}_{s,i}^{2}{c}_{s,i'}^{2}\left(\mathcal{H}_{1,0}\left({b}_{s,i}\right)-\mathcal{H}_{1,0}\left({b}_{s,i'}\right)\right)\right.\nonumber \\
 & +   \left({c}_{s,i}^{2}+{c}_{s,i'}^{2}\right)\left(\mathcal{H}_{1,1}\left({b}_{s,i}\right)-\mathcal{H}_{1,1}\left({b}_{s,i'}\right)\right) \nonumber\\
 & +   \left.\left(\mathcal{H}_{1,2}\left({b}_{s,i}\right)-\mathcal{H}_{1,2}\left({b}_{s,i'}\right)\right)\right]\!.\label{eq:DDCS_MY}
\end{align}
\end{widetext}
Note that for the Y model ($i=1$), the cross-terms vanish and expression~\eqref{eq:DDCS_MY} reduces to the much simpler
\allowdisplaybreaks{
\begin{align}
\frac{d^{2}\sigma_{Z_{s}}^{\text{Y}}}{d{k}d\Omega_{k}}  =&  {c}_{s}^{4}\mathcal{H}_{2,0}\left({b}_{s}\right)\\
&+2{c}_{s}^{2}\mathcal{H}_{2,1}\left({b}_{s}\right)+\mathcal{H}_{2,2}\left({b}_{s}\right).\label{eq:dsigma_photon_Y}
\end{align}}
The $\mathcal{H}_{l,m}$ integrals can be expressed in terms of the analytical results by Fronsdal and Überall \cite{Fronsdal58}, for which we use the notation from \cite{hau08}, $I_1$ and $I_2$:
\allowdisplaybreaks{
\begin{align}
\mathcal{H}_{1,0}({b}) &=\frac{I_2({b}^2)-I_2(0)+{b}^2}{{b}^4}\\
\mathcal{H}_{1,1}({b}) &=\frac{I_2(0)-I_2({b}^2)}{{b}^2}\\
\mathcal{H}_{1,2}({b}) &= I_2({b})\\
\mathcal{H}_{2,0}({b}) &=  2\frac{I_2({b}^2)-I_2(0)+{b}^2}{{b}^6} + \frac{I_1({b}^2)-1}{{b}^4}\\
\mathcal{H}_{2,1}({b}) &=\frac{I_2(0)-I_2({b}^2)}{{b}^4}-\frac{I_1({b}^2)}{{b}^2}\\
\mathcal{H}_{2,2}({b}) &= I_1({b}^2).
\end{align}}
The $I_1$ and $I_2$ integrals read
{\allowdisplaybreaks
\begin{align}
I_{1}({b}) &= \frac{\alpha Z^{2} r_{0}^{2}}{2\pi \,{k}\, {p}_{0}}\Bigg\{ 
16\,{p}\,(4{E}_{0}^{2}+{b}^{2})\frac{\|{\bm{p}}_{0}\times{\bm{k}}\|^{2}}{{k}^{2} W^{4}}\\
&\quad -\frac{2\,{p}}{W^{2}}\Big[\,4{E}_{0}^{2}+2{E}_{0}{E}-{E}\frac{D_{1}}{{k}}
+ {b}^{2}\big(1-2{E}\frac{{k}}{D_{1}}\big)\,\Big]\\
&\quad +\frac{2\,{p}}{W^{2}}\,
\frac{D_{1}^{2}+2({E}_{0}{E}-1)D_{1}+{b}^{2}(D_{1}-2{E}{k})}
{(D_{1}+{b}^{2})^{2}+4{p}^{2}{b}^{2}}\\
&\qquad\times\Bigg[\,\frac{1}{D_{1}}\big(16{E}_{0}{E}-4{E}_{0}^{2}{b}^{2}-b^{4}\big)
-\frac{4{E}_{0}^{2}+{b}^{2}}{{k}^{2}W^{2}}\,\\
&\qquad \;\times\{D_{1}^{2}+2({E}_{0}{E}-1)D_{1}+{b}^{2}(D_{1}-2{E}{k})\}\Bigg]\\
&\quad -\frac{2{k}^{2}{p}}{(D_{1}+{b}^{2})^{2}+4{p}^{2}{b}^{2}}
\Bigg[\frac{4}{D_{1}^{2}}\big(4{E}_{0}^{2}+(1-D_{1}){b}^{2}\big)\\
&\qquad +\frac{D_{1}^{2}+2({E}_{0}{E}-1)D_{1}+{b}^{2}(D_{1}-2{E}{k})}
{D_{1}\,\|\,{\bm{p}}_0-{\bm{k}}\,\|^{2}}\Bigg]\\
&\quad -\frac{4{k}}{D_{1}}\ln({E}+{p})
+\frac{L_{1}}{W}\Bigg[\,2{k}+\frac{4{k}}{D_{1}}({E}_{0}^{2}+{p}^{2}+{b}^{2})\,\\
&\quad +\frac{2}{D_{1}W^{2}}({E}_{0}D_{1}-2{k}+{b}^{2}{k})\\
&\qquad\times\Big(8{E}_{0}{E}-\frac{D_{1}^{2}}{2}-{b}^{2}(2{E}_{0}^{2}+2{p}^{2}+D_{1})-b^{4}\Big)\\
&\quad +\frac{2(2{E}_{0}^{2}+{b}^{2})(D_{1}-2{E}{k})+D_{1}^{2}+2({E}_{0}{E}-1)D_{1}}
{{k}\,W^{2}}\\
&\quad -\frac{3}{{k}\,W^{4}}(4{E}_{0}^{2}+{b}^{2})\Big({E}_{0}\frac{D_{1}}{{k}}-2+{b}^{2}\Big)
\\
&\qquad\big(D_{1}^{2}+2({E}_{0}{E}-1)D_{1}+{b}^{2}(D_{1}-2{E}{k})\big)\Bigg]\\
& +\frac{{k}^{2}L_{2}}{D_{1}\,\|\,{\bm{p}}_0-{\bm{k}}\,\|}
\Big[\,\frac{2}{D_{1}}-2+\frac{D_{1}-2{E}{k}}{2\,\|\,{\bm{p}}_0-{\bm{k}}\,\|^{2}}\,\Big]
\Bigg\}\,, 
\end{align}}
and
{\allowdisplaybreaks
\begin{align}
I_{2}({b}) &= - \frac{\alpha Z^{2} r_{0}^{2}}{2\pi {k}\,{{p}_{0}}}
\Biggl\{  \frac{2 {p}\left( 4 {E}_{0}^{2} + {b}^{2} \right)}{W^{2}} 
\left( {E}_{0} \frac{D_{1}}{{k}} - 2 + {b}^{2} \right) \nonumber \\
&+ \frac{L_{1}}{W} \Bigl[ {k}(D_{1} + 2{b}^{2})\\ \nonumber
&\quad+ \frac{2{k}}{D_{1}} \bigl({b}^{4} + 2{b}^{2}({E}_{0}^{2} + {p}^{2}) 
- 8 {E}_{0}{E} \bigr) \Bigr] \nonumber \\
&+ \frac{4 {E}_{1}^{2} + {b}^{2}}{{k} W^{2}}
\Bigl( D_{1}^{2} + 2({E}_{0}{E} - 1)D_{1} \\
&\quad+ {b}^{2}(D_{1} - 2{E}{k}) \Bigr) \nonumber \\
&+ \frac{{k}^{2} L_{2}}{D_{1} \, \|{\bm{p}_{0}} - {\bm{k}}\|}
\Biggl[ \frac{2}{D_{1}} \bigl(4 {E}^{2} + {b}^{2}(1 - D_{1})\bigr) \nonumber \\
&\quad + \frac{D_{1}^{2} + 2({E}_{0}{E} - 1)D_{1} 
+ {b}^{2}(D_{1} - 2{E}{k})}{2\|{\bm{p}_{0}} - {\bm{k}}\|^{2}} \Biggr] \nonumber \\
&- \frac{4{k}}{D_{1}} \,{b}^{2} \ln({E} + {p})
\Biggr\}.
\end{align}}
The quantities involved in $I_1$ and $I_2$ are given by 
\begin{align}
D_{1} &= 2{k}\,\bigl({E}_{0} - {{p}_{0}}\cos\theta_{0}\bigr)\\
W &= \Biggl[\left({{p}_{0}}\,\frac{D_{1}}{{k}}\right)^{2}
      + 2{b}^{2}\left({E}_{0}\frac{D_{1}}{{k}} - 2\right)
      + {b}^{4}\Biggr]^{1/2}\\
L_{1} &= \ln \frac{({E}_{0}{E}-1)\,D_{1}/{k} + {E}\,{b}^{2} + {p}\,W}%
                  {({E}_{0}{E}-1)\,D_{1}/{k} + {E}\,{b}^{2} - {p}\,W}\\
L_{2} &= \ln \frac{(\bigl\|{\bm{p}_{0}} - {\bm{k}}\bigr\| + {p})^{2} + {b}^{2}}%
                  {(\bigl\|{\bm{p}_{0}} - {\bm{k}}\bigr\| - {p})^{2} + {b}^{2}}.
\end{align}
Equation~\eqref{eq:DDCS_MY} is a \emph{fully analytical} expression of the screened doubly differential cross section (DDCS) for \emph{arbitrary ionized states} described by a multi-Yukawa (MY) potential. That result, together with Eq.~\eqref{eq:cross-section=0000202BN-1} and the numerical integration of Eq.~\eqref{eq:RDP_tot}, enables the evaluation of the total DDCS Eq.~\eqref{eq:OMW}.

\section{Cross Section Results} \label{sec:Results}
The equations~\eqref{eq:cross-section=0000202BN-1},~\eqref{eq:RDP_tot}, and~\eqref{eq:DDCS_MY} have been implemented in a \textsc{Matlab} program~\cite{brem_cross_sec_release}. While Eqs.~\eqref{eq:cross-section=0000202BN-1} and~\eqref{eq:DDCS_MY} can be evaluated directly from their analytical forms, the RDP contribution is evaluated by numerically integrating the TDCS in Eq.~\eqref{eq:RDP_tot} over the scattering angles $(\theta,\phi)$ using an adaptive grid method, achieving a relative accuracy better than $10^{-6}$~\cite{brem_cross_sec_release}.
\begin{figure}[hbtp!]
\centering
\includegraphics[width=0.49\textwidth]{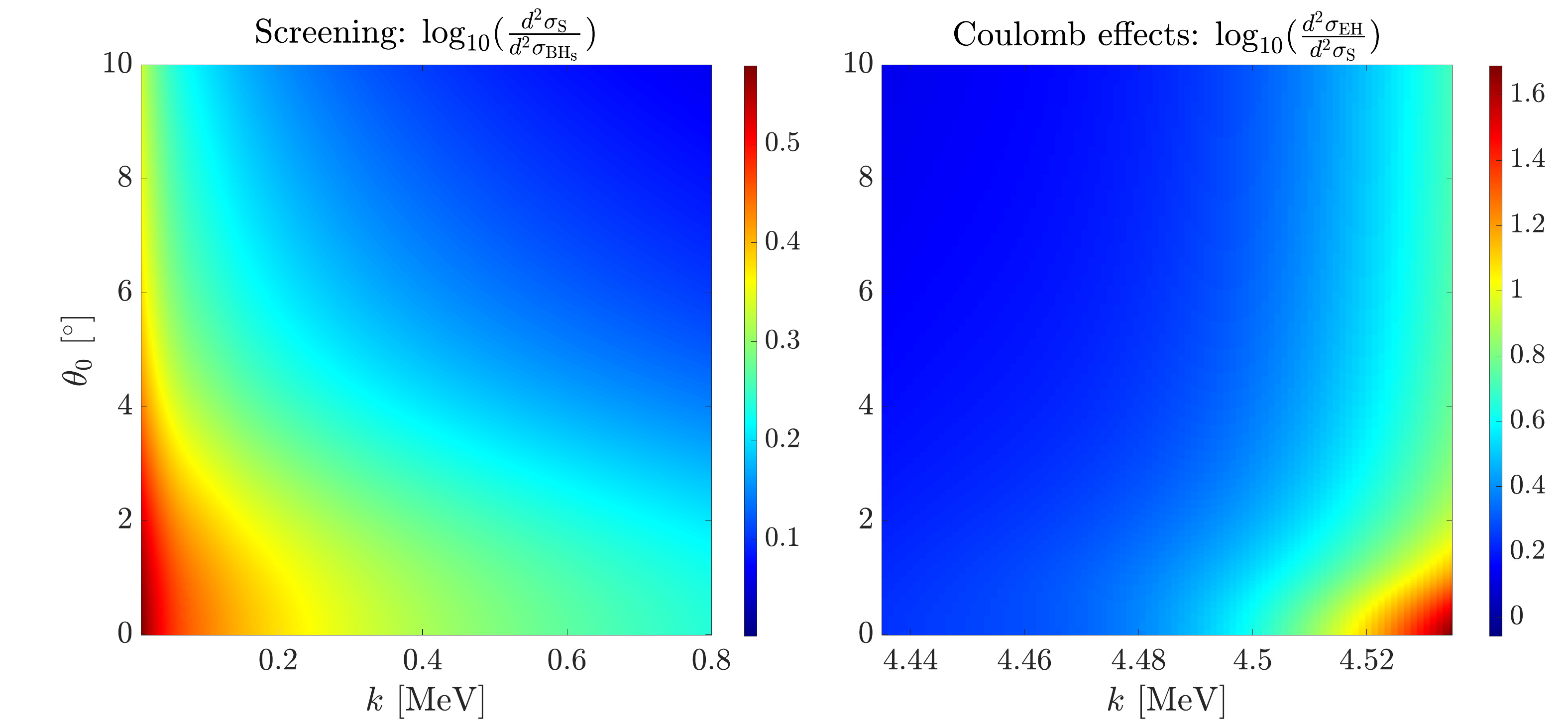}
\caption{Left: Region of the parameter space where the screening is prominent, namely soft-photons and forward emissions, for an incoming electron of energy $E_0=4.54$ MeV on a gold target. Right: Region of the parameter space where Coulomb corrections are required, i.e. hard photons and forward emissions. In the ratios, $d^2\sigma$ stands for $d^2\sigma/dkd\Omega_k$.} 
\label{fig:physical_effects}
\end{figure}

Figure~\ref{fig:physical_effects} provides physical motivation for the separation of screening and Coulomb contributions embodied in the Olsen--Maximon--Wergeland (OMW) formulation, Eq.~\eqref{eq:OMW}. The left panel shows the logarithm of the ratio between the unscreened (Sauter) reference and the corresponding screened Bethe--Heitler (multi-Yukawa) DDCS, as a function of the emitted photon energy and emission angle, for an incident electron energy of 
$E_0=4.54$ MeV.
 This representation highlights the importance of screening effects, which are most pronounced in the soft-photon, forward-emission domain. The right panel displays the logarithm of the ratio of the leading-order RDP (Elwert--Haug) cross section to Sauter’s Coulomb-field result, illustrating the onset of Coulomb (nuclear) corrections toward forward angles and higher photon energies. Because the next-to-leading-order RDP expression is not reliable in the high-frequency (tip) region, the leading-order result is used in the ratio shown.\footnote{The next-to-leading-order RDP theory was derived under the assumption that both the incident and scattered electron energies are large, and therefore loses accuracy in the high-frequency (tip) region.} 
 
Figure~\ref{fig:physical_effects} is intended to illustrate the physical motivation for treating the screening correction additively in Eq.~\eqref{eq:OMW}; a detailed assessment of the OMW additivity rule can be found in Ref.~\cite{JakubassaMangia_2019}.

\begin{figure}[b!]
    \centering
    \includegraphics[width=1\linewidth]{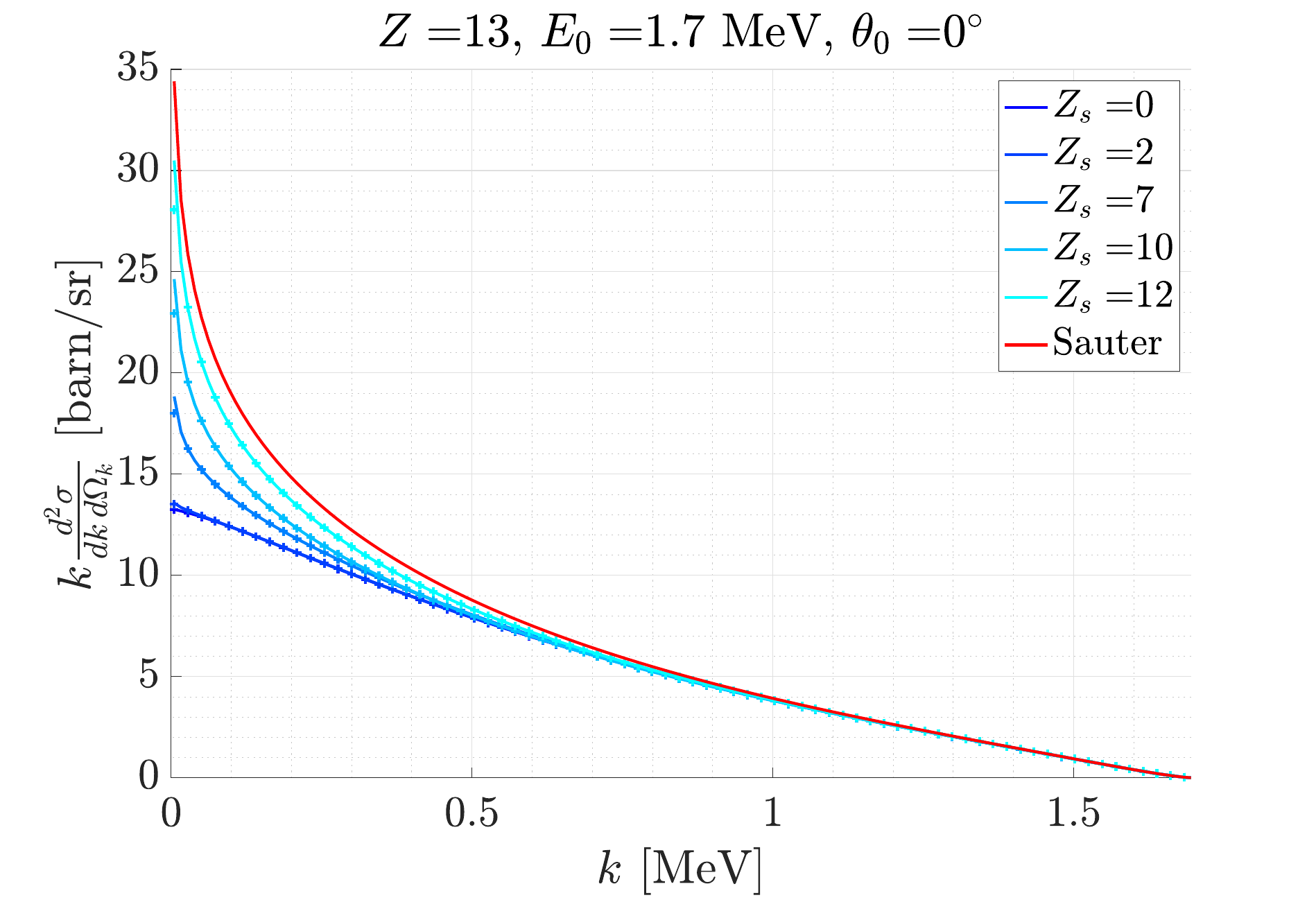}
       \caption{Comparison of the DDCS obtained from the closed form Eq.~\eqref{eq:DDCS_MY} (solid lines) and from a numerical integration of Eq.~\eqref{eq:dsigmabrem3} (markers), for several ionized states of aluminum, and two Yukawa exponentials. }\label{fig:AnVsNum}
\end{figure}
To assess the consistency of the analytical DDCS for partially ionized species, results from the closed-form expression~\eqref{eq:DDCS_MY} are compared with those obtained by direct numerical integration of Eq.~\eqref{eq:dsigmabrem3} for all ionization states of a given atom. Fig.~\ref{fig:AnVsNum} shows an excellent agreement for several ionization degrees of aluminum ($Z=13$), where two Yukawa exponentials have been used for the description of the potential. As expected, in the screening-dominant (large wavelength) region, the emission increases with the ionization state, i.e. with the effective ion charge $Z_{s}$. The fully stripped limit is Sauter's expression Eq.~\eqref{eq:cross-section=0000202BN-1} for a pure Coulomb field. The neutral case $Z_0$ corresponds by definition to the expression found by Haug~\cite{hau08}, except that in the present model the number of exponentials is $Z$-dependent and not fixed to three, allowing a better fitting of the atomic potential. 
\begin{figure}[b!]
    \centering
    \includegraphics[width=1\linewidth]{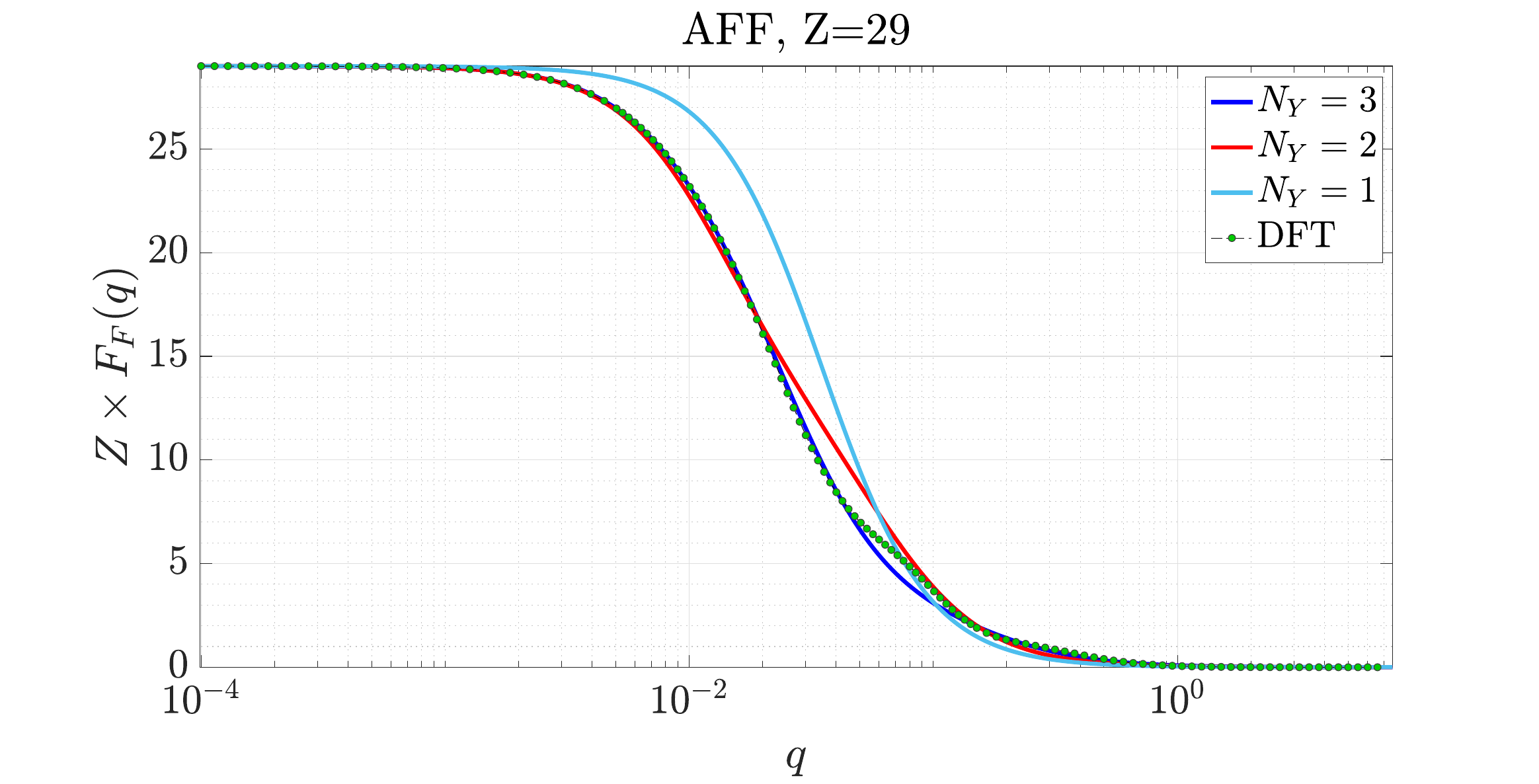}\\[1ex]
    \includegraphics[width=1\linewidth]{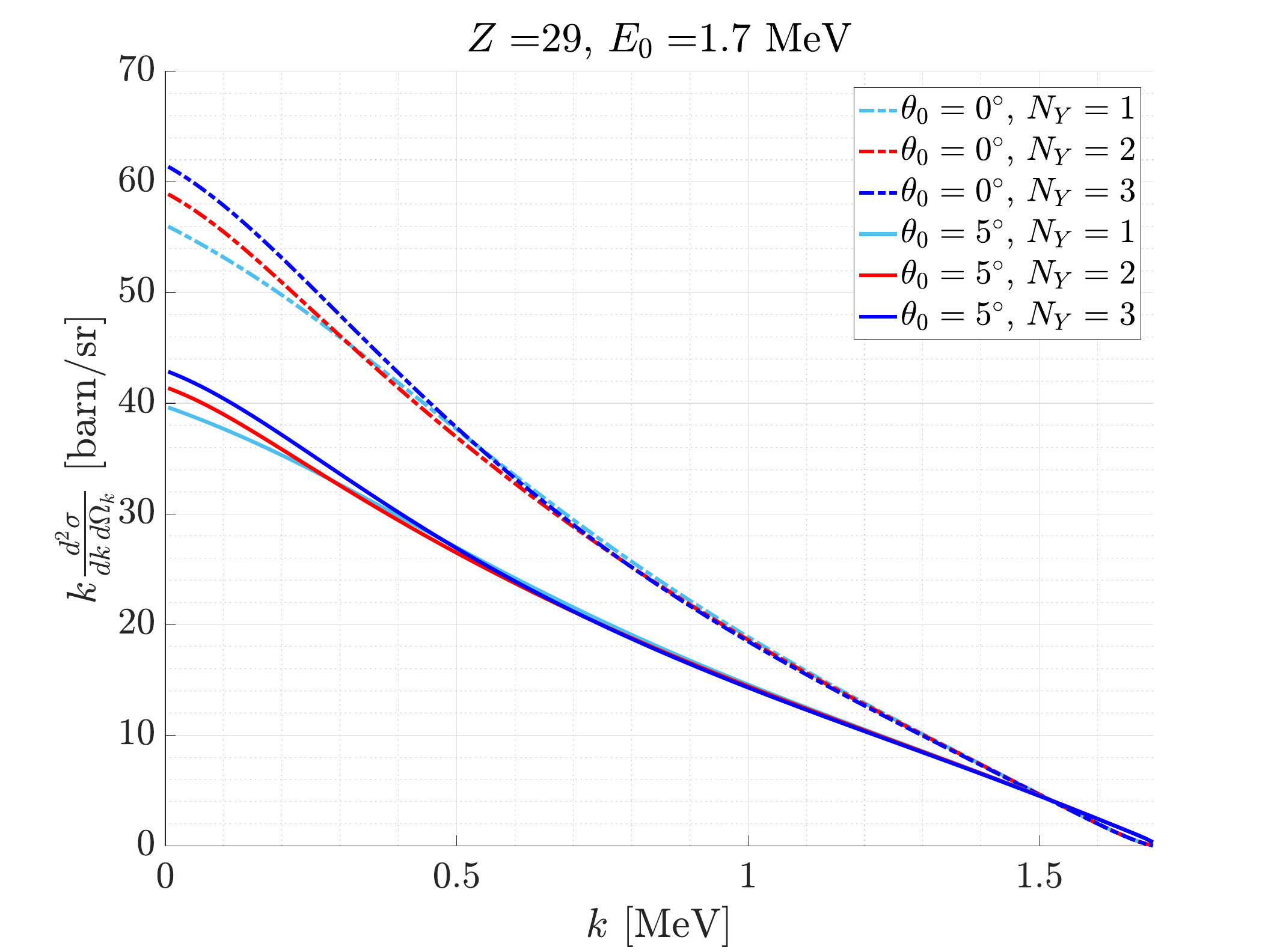}
       \caption{Top: Atomic form factor for a neutral copper atom ($Z=29$) as a function of the recoil momentum, calculated with one, two, and three Yukawa exponentials, and compared with the reference form factor obtained from DFT. Bottom: Neutral doubly differential cross section (DDCS), normalized by the inverse photon energy, for one, two, and three Yukawa exponentials and for two emission angles.} \label{fig:AFFandMY}
\end{figure}

\subsection{Multi-Yukawa and Ionized Cross Sections}\label{sec:Multiyuk}
This section analyzes the influence of the multi-Yukawa potential representation on the bremsstrahlung emission cross sections of partially ionized species, highlighting the model’s internal consistency and physical trends.

We begin by examining how the atomic form factor (AFF) is influenced by the number of exponentials, $N_Y$, used in the multi-Yukawa model. The top panel of Fig.~\ref{fig:AFFandMY} shows the AFF as a function of the nuclear recoil momentum for a neutral copper atom, calculated with several values of $N_Y$ and compared with the “exact” form factor obtained from density-functional theory (DFT)~\cite{Walkowiak22,sav23} using the \textsc{Gaussian} code~\cite{GAUSSIAN}. The bottom panel of Fig.~\ref{fig:AFFandMY} illustrates the corresponding impact on the doubly differential cross section (DDCS). The incident-electron energy is set to 1.7 MeV, and two small emission angles, $0^{\circ}$ and $5^{\circ}$, are considered, since screening effects are most pronounced in the forward direction. As expected, the DDCS increases slightly with $N_Y$, because the screening factor for $N_Y=1$ is larger than for $N_Y=2,3$ at low values of $q$.

\begin{figure}[hbtp!]
    \centering
    \includegraphics[width=1\linewidth]{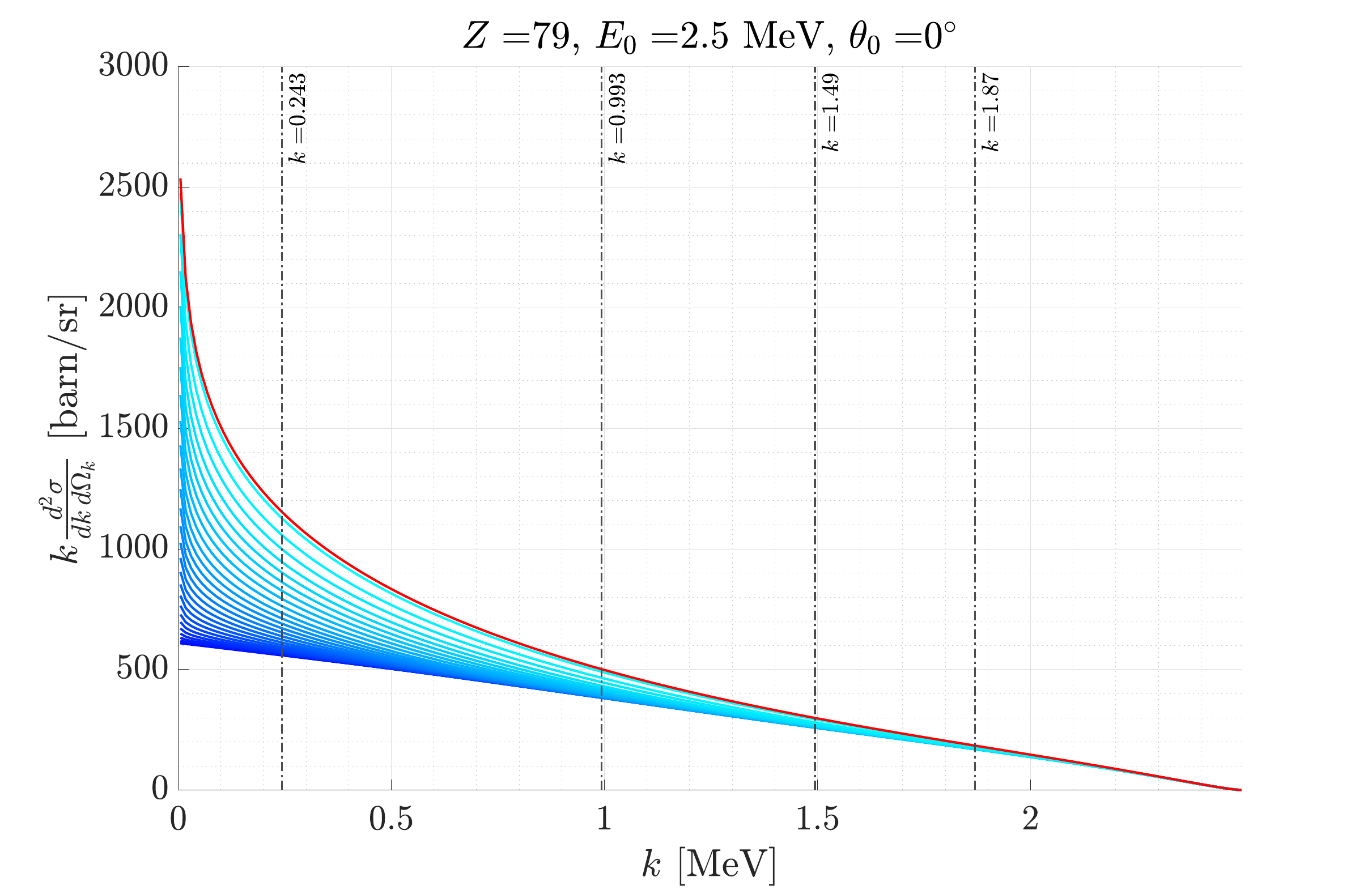}\\[1ex]
    \includegraphics[width=1\linewidth]{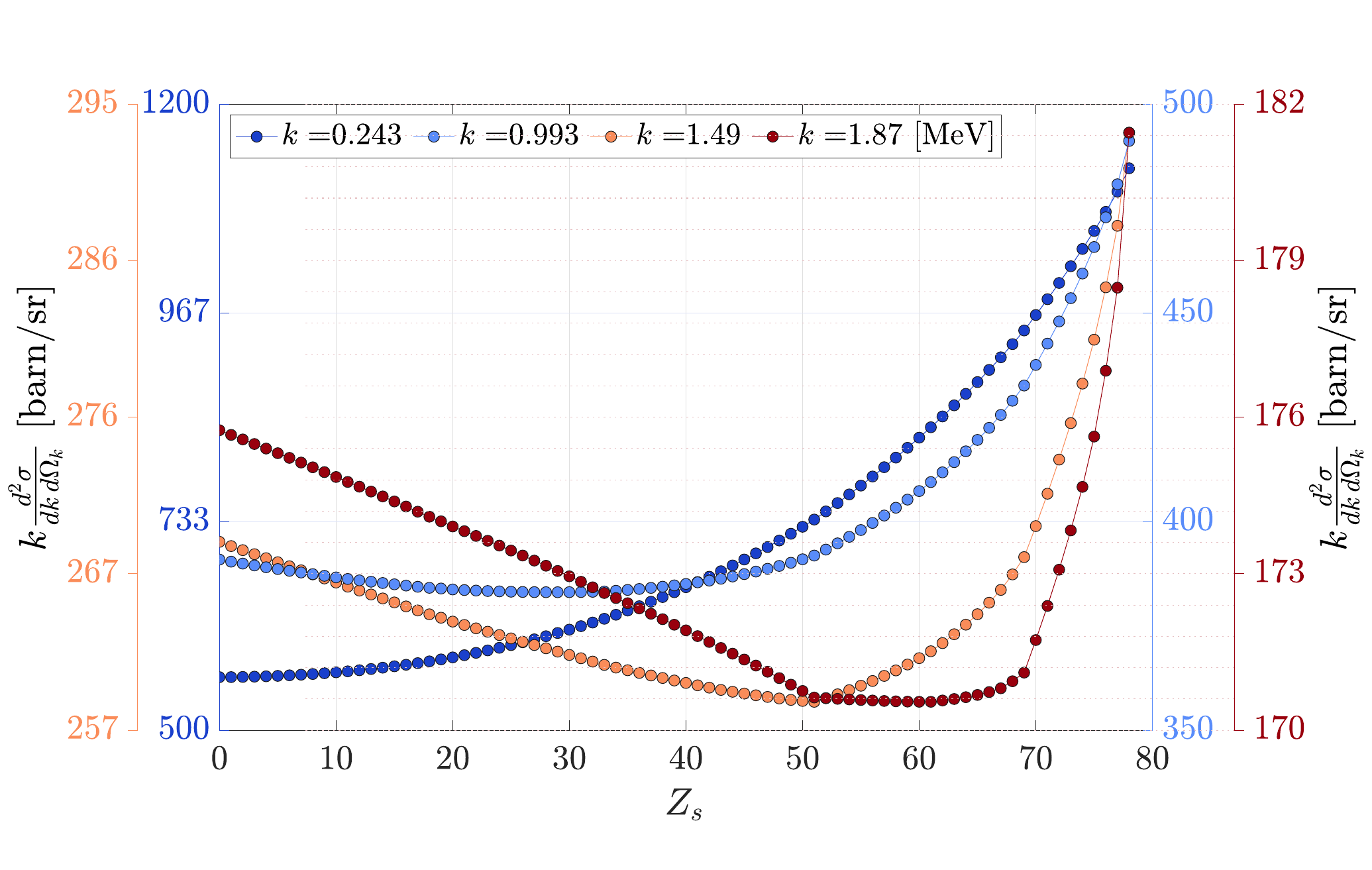}
      \caption{Top: Ionized doubly differential cross section (DDCS) for gold. Every third ionization level is shown; the red line represents Sauter’s model. Vertical lines indicate the photon energies for which the DDCS is plotted as a function of ionization level in the bottom inset. Bottom: DDCS for gold at the four photon energies indicated above.}\label{fig:Ionized_Gold}
\end{figure}

The DDCS is calculated for various ionized states of  gold ($Z=79$) and is shown in Fig.~\ref{fig:Ionized_Gold}. Although the cross section is monotonically increasing with the ion charge $Z_{s}$ for low values of $k$, a minimum is found at non-neutral states for higher photon energies.

\begin{figure}[hbtp!]
    \centering
    \includegraphics[width=1\linewidth]{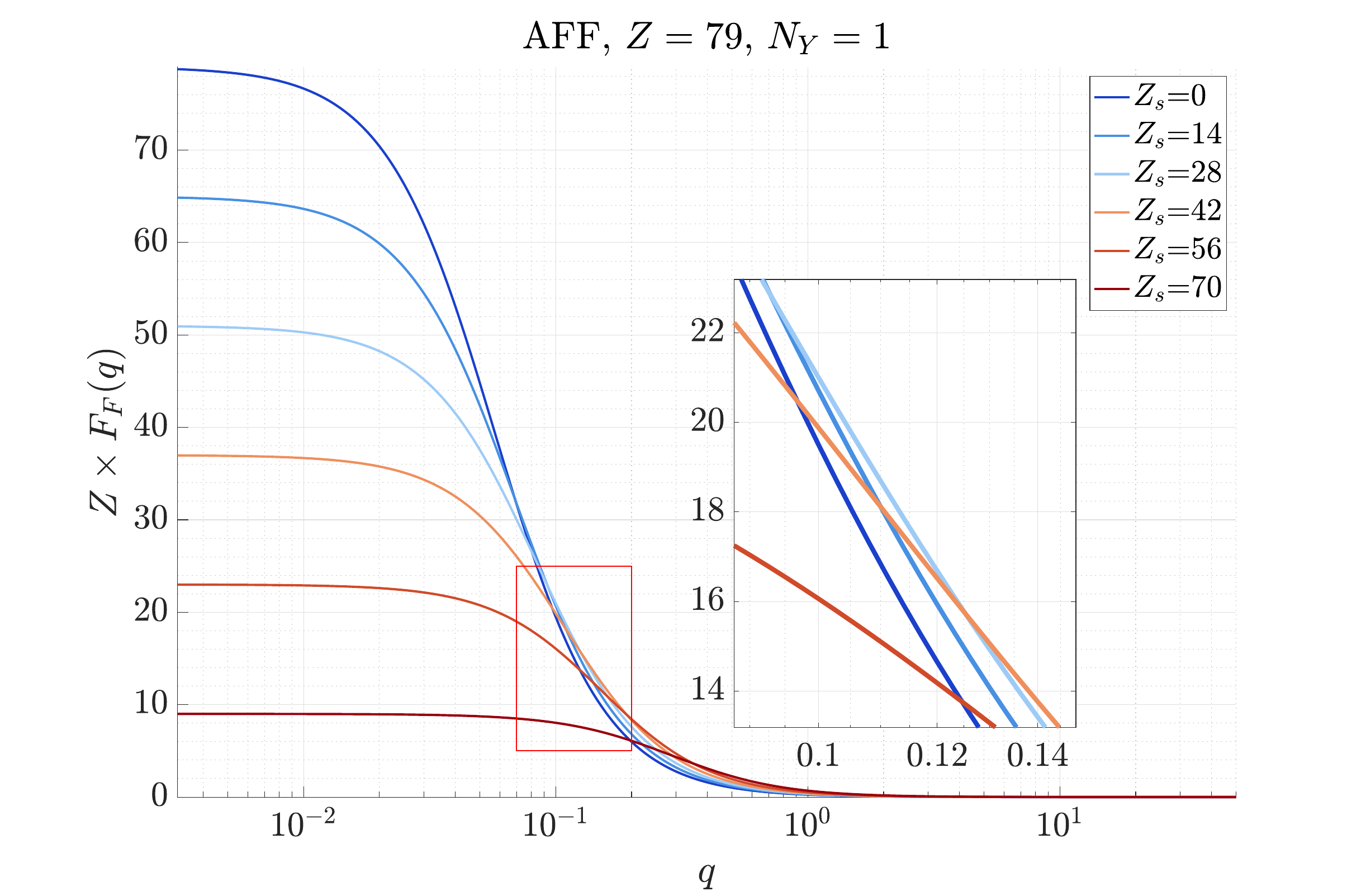}
       \caption{Atomic form factor for several ionization states of gold, and one Yukawa exponential. The form factor exhibits some crossings of the ionized curves, as emphasized by a zoom in the red rectangle.} \label{fig:AFF_zoomed_crossing}
\end{figure}

The non-monotonic behavior of the doubly differential cross section with increasing ionization state at higher photon energies $k$ originates from the atomic form factor, which reflects the effective nuclear charge experienced during the interaction. At higher photon energies, the form factor does not vary monotonically with $Z_{s}$, as illustrated in Fig.~\ref{fig:AFF_zoomed_crossing}, where $F_{Z_{s}}$ is plotted as a function of the recoil momentum $q(k)$ for several ionization states.

Since the atomic form factor is the Fourier transform of the spatial electronic density $n_{s}$,
\begin{align}
F_{Z_{s}}({q}) &= \frac{1}{Z} \int_{\mathbb{R}^3} 
n_{s}(\mathbf{r}) \, e^{i{\bm{q}}\cdot\mathbf{r}} \, d^3r \\
&= 1 - \sum_i {A}_{s,i} \, 
\frac{{q}^2 + q_{Z_{s}}{b}_i^2}{{q}^2 + {b}_i^2},
\label{eq:AFF_Fourier}
\end{align}
where $\boldsymbol{r}$ is in units of the reduced Compton wavelength and $n_s$ is rescaled accordingly so as to preserve its normalization, the relative behavior of the form factors for different 
ionization states is strongly influenced by the corresponding radial 
electron densities around the nucleus. When these densities cross, the form 
factors can exhibit similar non-monotonic behavior.

\begin{figure}[hbpt!]
    \centering
    \includegraphics[width=0.49\textwidth]{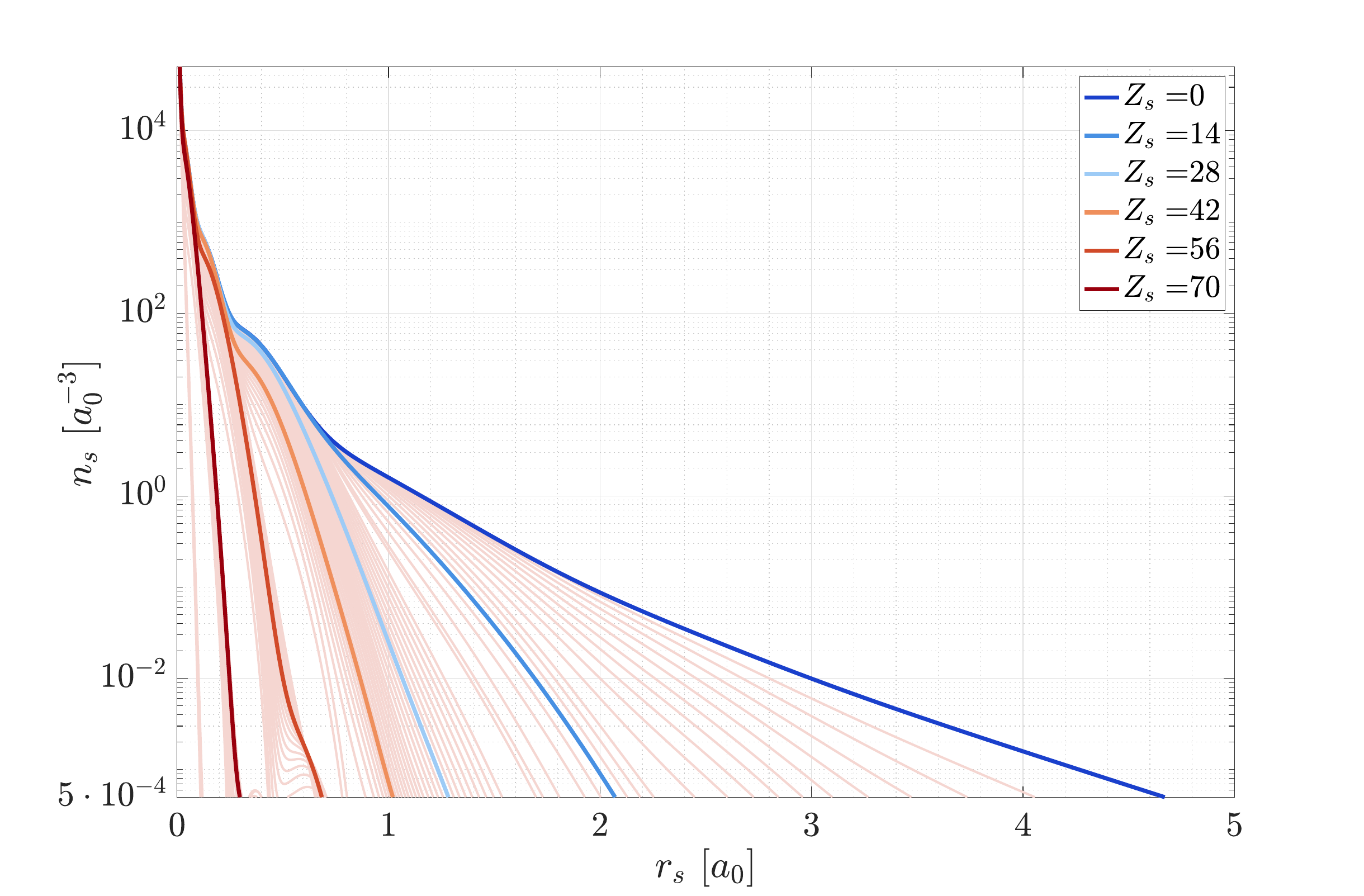}\\[1ex]
    
    \includegraphics[width=0.49\textwidth]{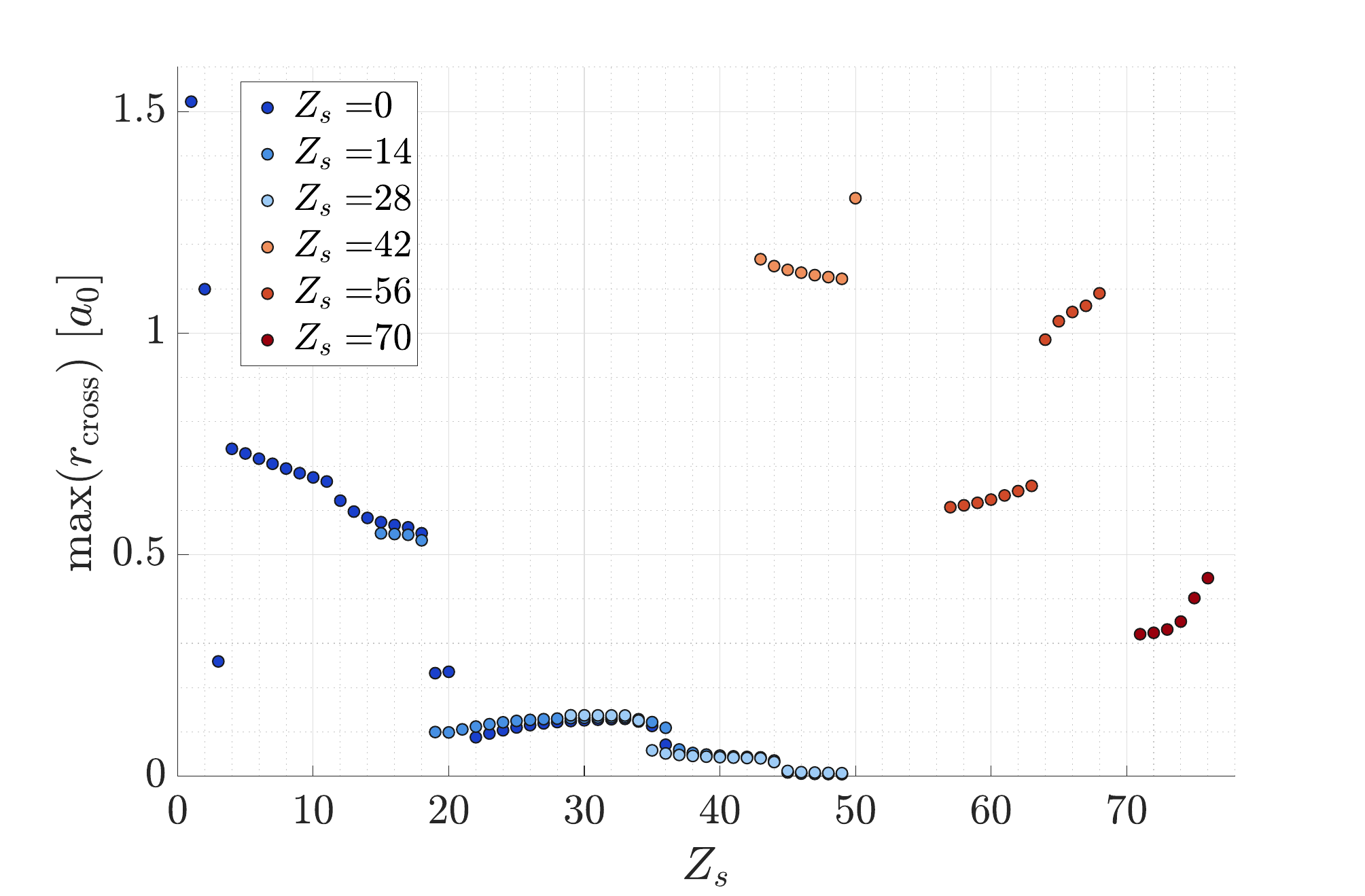}
       \caption{
Top: Bound-electron density profiles for all ionization states of gold, calculated using DFT with the \textsc{Gaussian} code~\cite{GAUSSIAN}. 
The upper blue curve corresponds to the neutral atom and the ionization degree increases from right to left. 
Bottom: Largest intersection radius between each reference density 
(highlighted in the legend) and all higher ionization states for a gold atom.
}\label{fig:DFT_Z79}
\end{figure}

For each ionization state of gold shown in Fig.~\ref{fig:AFF_zoomed_crossing}, the corresponding radial bound-electron density is displayed in the top inset of Fig.~\ref{fig:DFT_Z79}. Because the crossings between the density profiles are not clearly visible, the largest intersection radius between each reference profile and  higher ionized states has been determined. The resulting characteristic intersection radii are presented in the bottom panel of Fig.~\ref{fig:DFT_Z79}.

The intersection radius between two radial bound-electron density profiles, 
$n_{s}(r)$ and $n_{s'}(r)$, provides a convenient way to visualize how ionization 
modifies the spatial extent of the electronic cloud. 
If the bound-electron distribution contracted strictly monotonically with increasing 
ionization, no radial crossings would occur. Each more-ionized state would be confined 
to a smaller radius. 
Conversely, a radial crossing between $n_{s}$ and $n_{s'}$ indicates that upon 
removing electrons, the residual bound charge has redistributed onto a larger 
characteristic shell at a certain radius, so that the screening length is 
increased rather than decreased. 
Among the possible intersections, the outermost crossing is the most physically 
relevant, as it occurs in the outer region of the atom where the residual bound-electron 
density dominates and thus determines the long-range screening behavior (low $q$-domain).
The presence of such crossings therefore indicates that the contraction of the electronic 
cloud with ionization is not perfectly monotonic: the residual electrons undergo 
redistribution which in turn leads to the non-monotonic behavior of the atomic form factor $F_{Z_{s}}(q)$ with 
ionization, as observed in Fig.~\ref{fig:AFF_zoomed_crossing}.

The existence of non-monotonic behaviors in the screened cross section illustrated in Fig.~\ref{fig:Ionized_Gold}, is systematically observed across different incident electron energies, emission angles, and atomic numbers. Since the formulation is constrained by fully relativistic ab initio density functional theory calculations, the resulting electronic densities can be regarded as essentially exact.
The observed effect therefore reflects intrinsic atomic-physics behavior rather than a model-dependent artifact.
Whereas this feature does not appear in approximate models such as the Tseng--Pratt--Avdonina--Lamoureux formulation, it is found in several alternative descriptions that incorporate comparable physical ingredients, including the Thomas--Fermi~\cite{Thomas_1927, Fermi_28}, Thomas--Fermi--Kirillov~\cite{kir75}, and Tseng--Pratt--Botto~\cite{Botto_78, Walkowiak22} models.
A detailed comparison among these screening formalisms lies however beyond the scope of the present work.

\subsection{Comparison with Data for the Doubly Differential Cross Section}
\begin{figure*}[!t]
    \centering
    \includegraphics[width=0.49\linewidth]{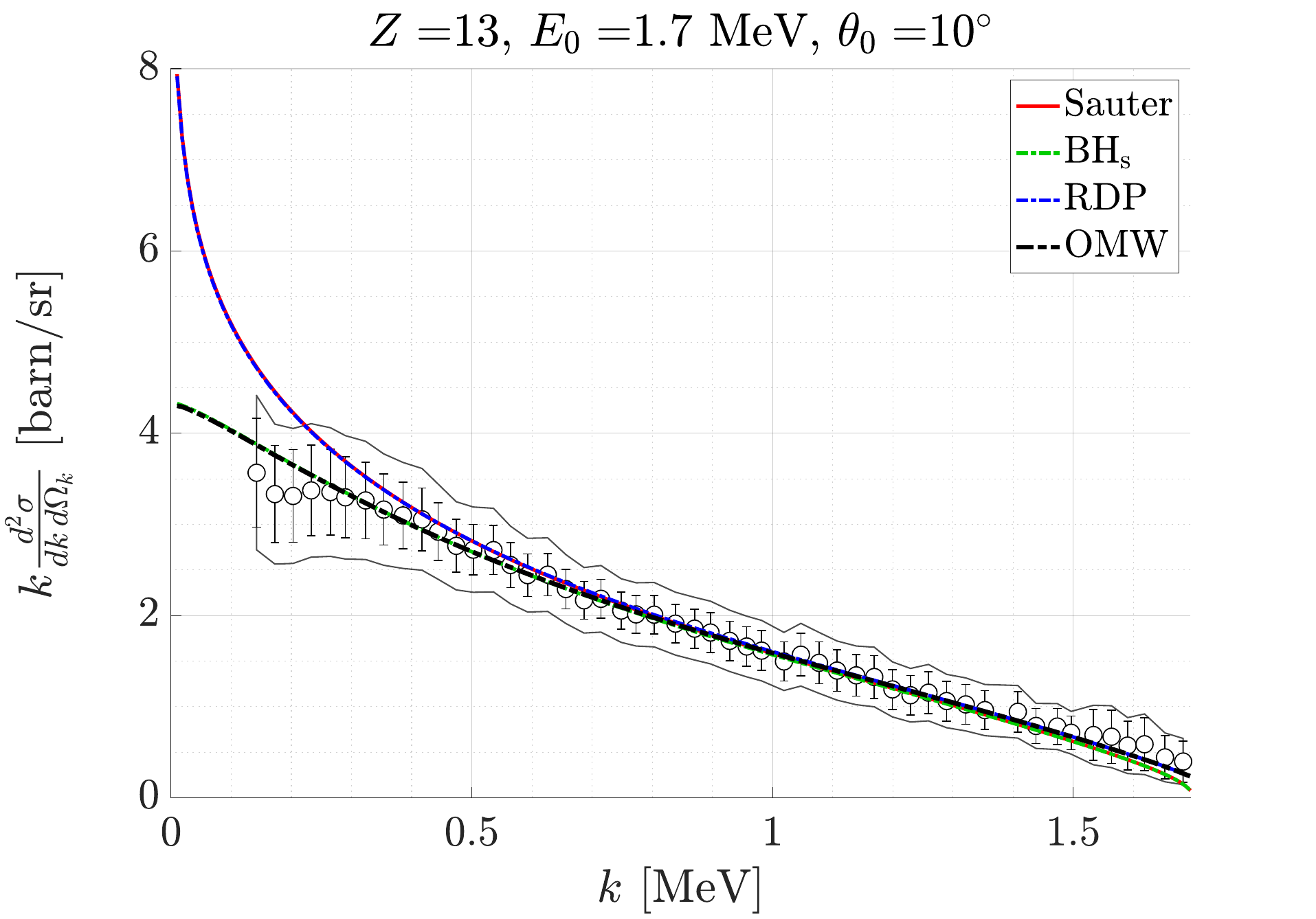}
    \includegraphics[width=0.49\linewidth]{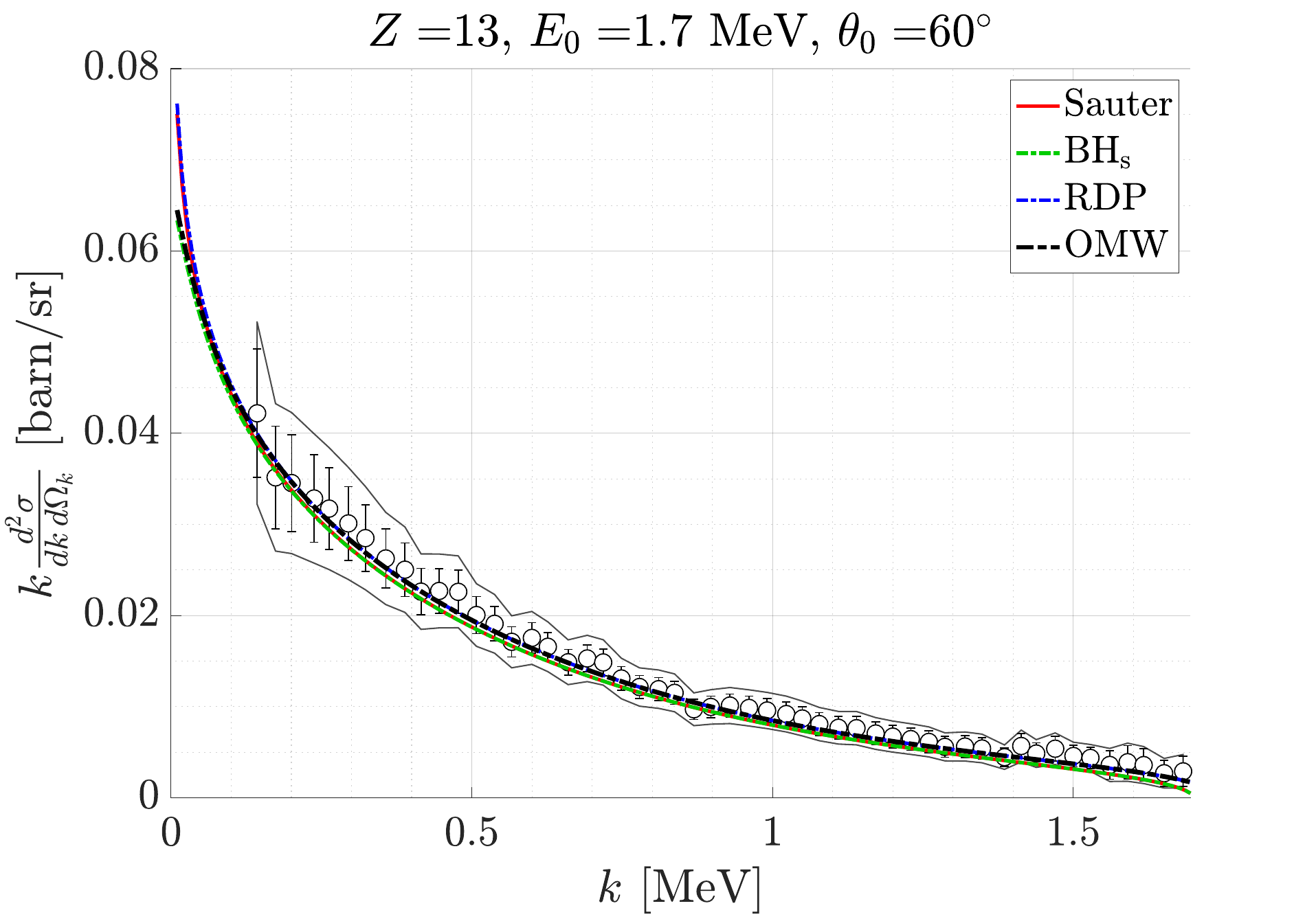}\\[1ex]
    
    \includegraphics[width=0.49\linewidth]{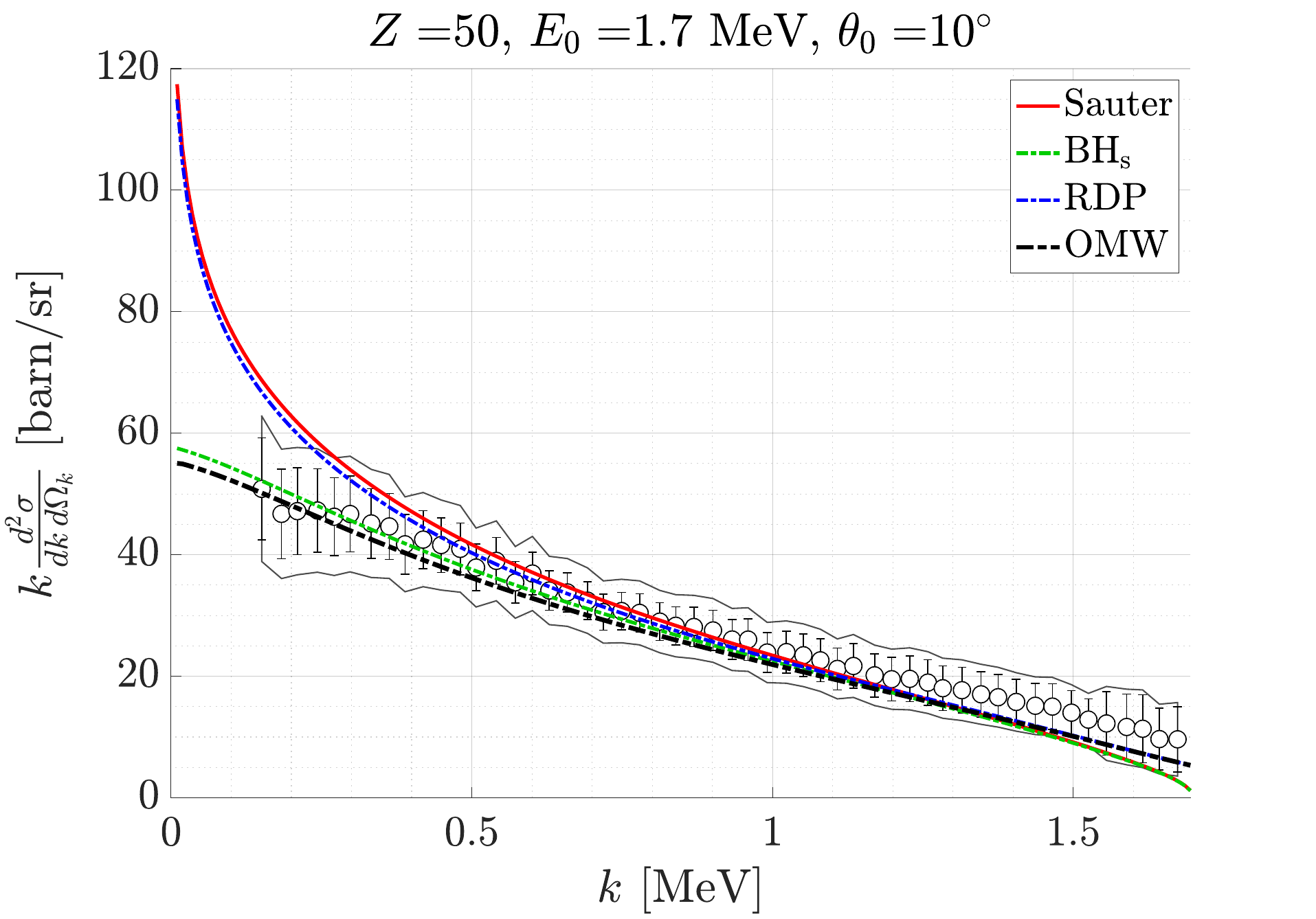}
    \includegraphics[width=0.49\linewidth]{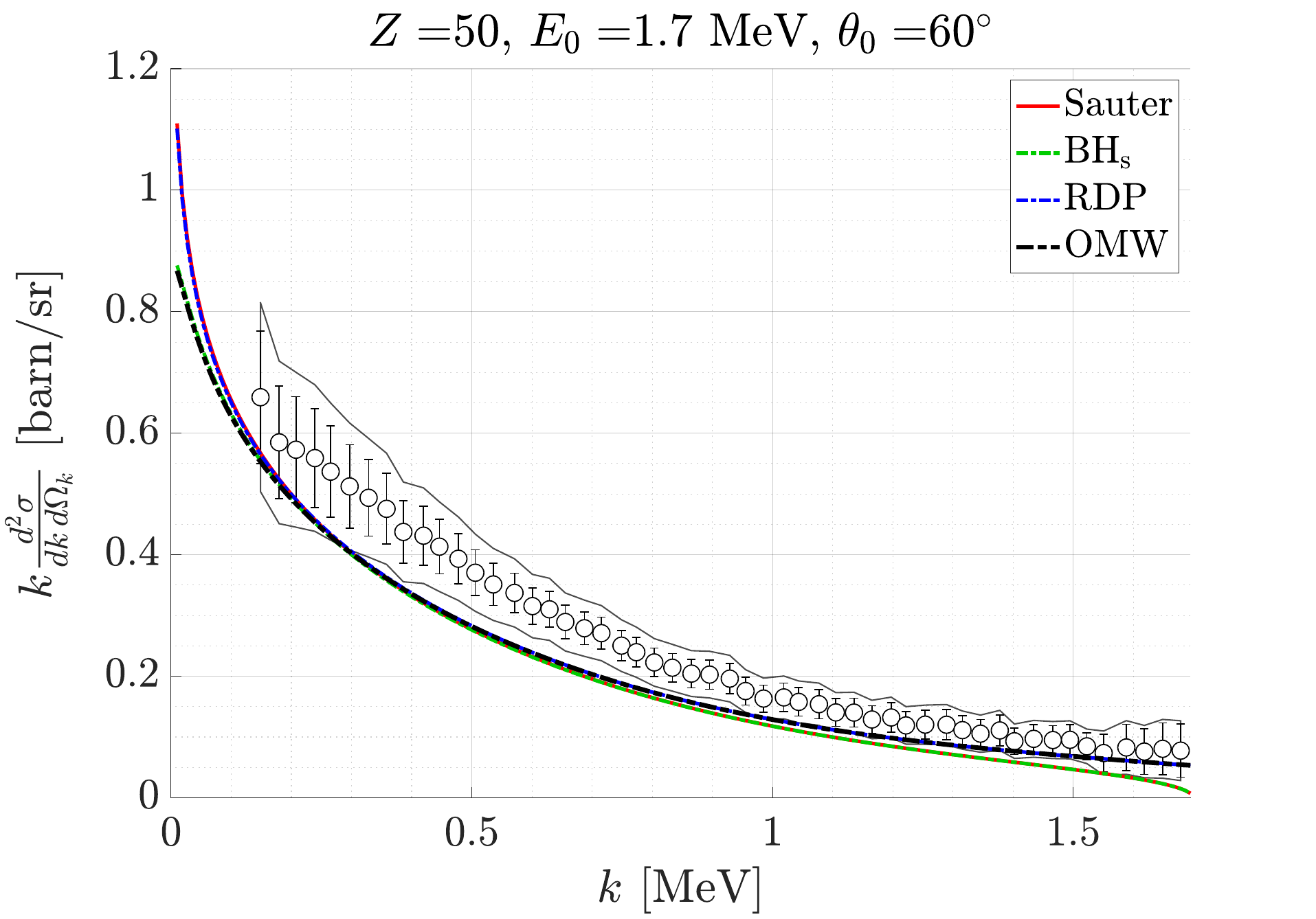}\\[1ex]
    
    \includegraphics[width=0.49\linewidth]{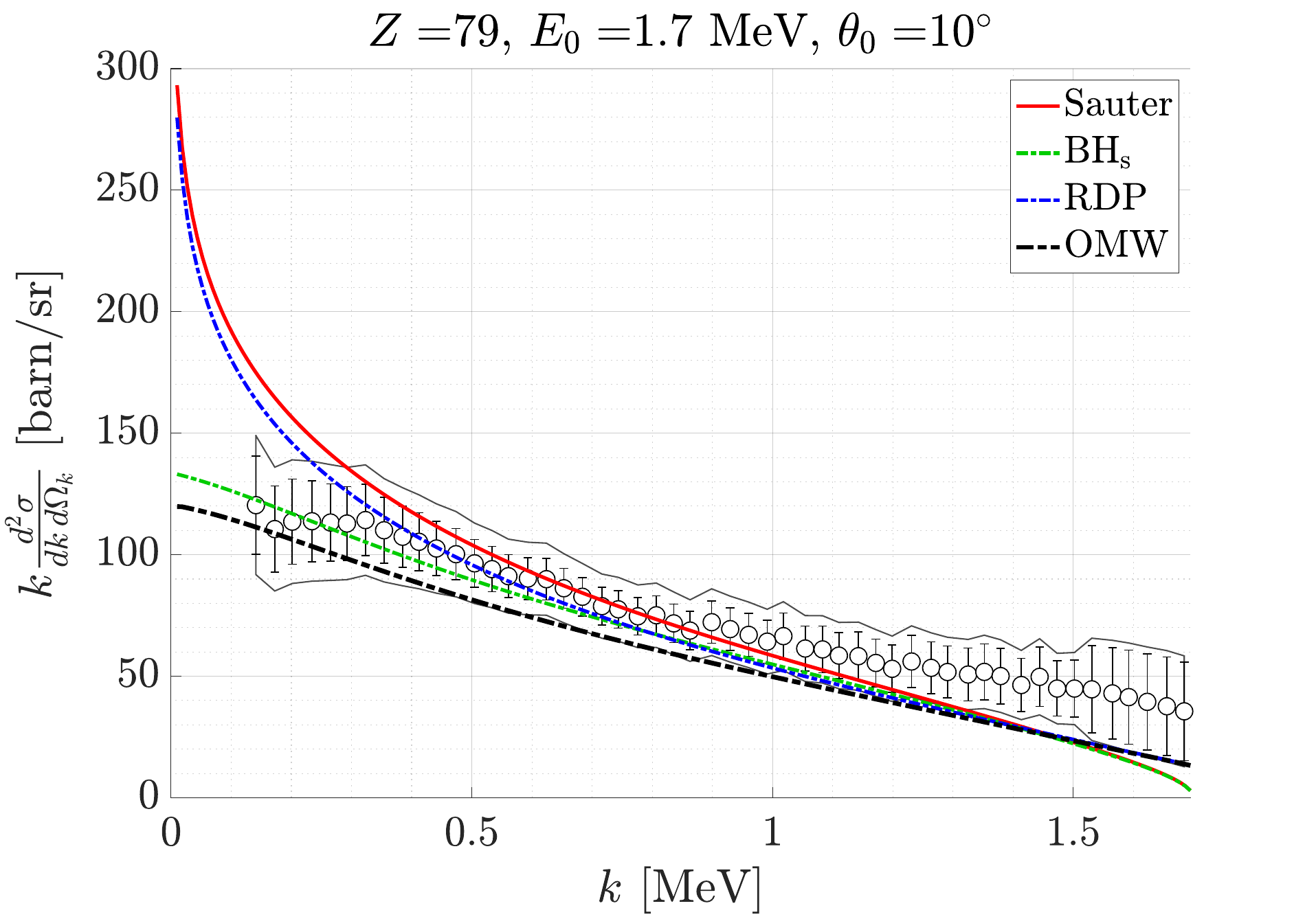}
    \includegraphics[width=0.49\linewidth]{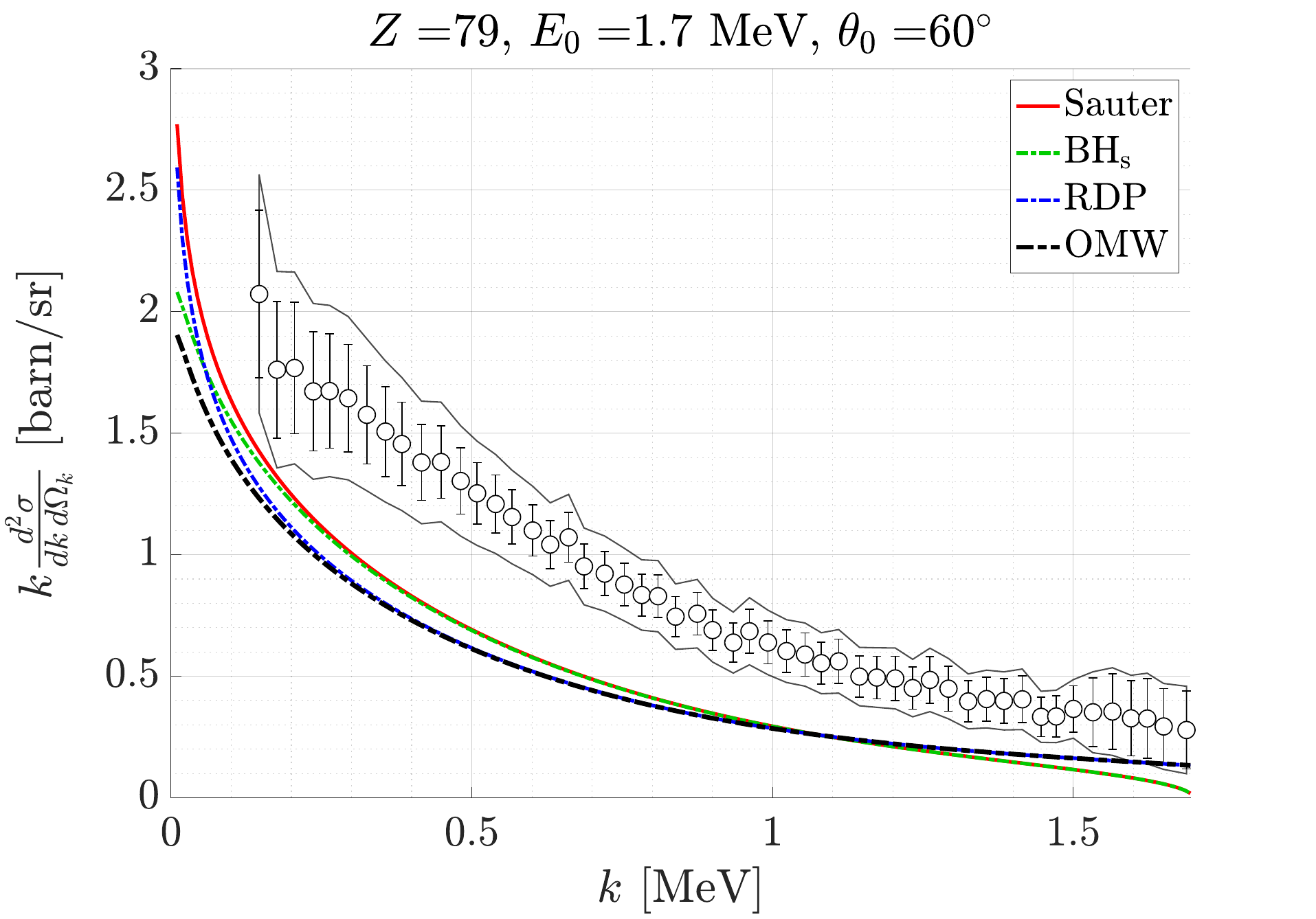}
    \caption{Comparison of the theoretical model components against the experimental results by Rester and Dance \cite{Rester67}, for aluminum, tin and gold. The error bars represent statistical uncertainties, while the contours indicate the sum of the statistical and systematic errors. The incoming electron has an energy of 1.7 MeV.}
\label{fig:Z_theta0}
\end{figure*}
\begin{figure*}[!t]
    \centering
    \includegraphics[width=0.49\linewidth]{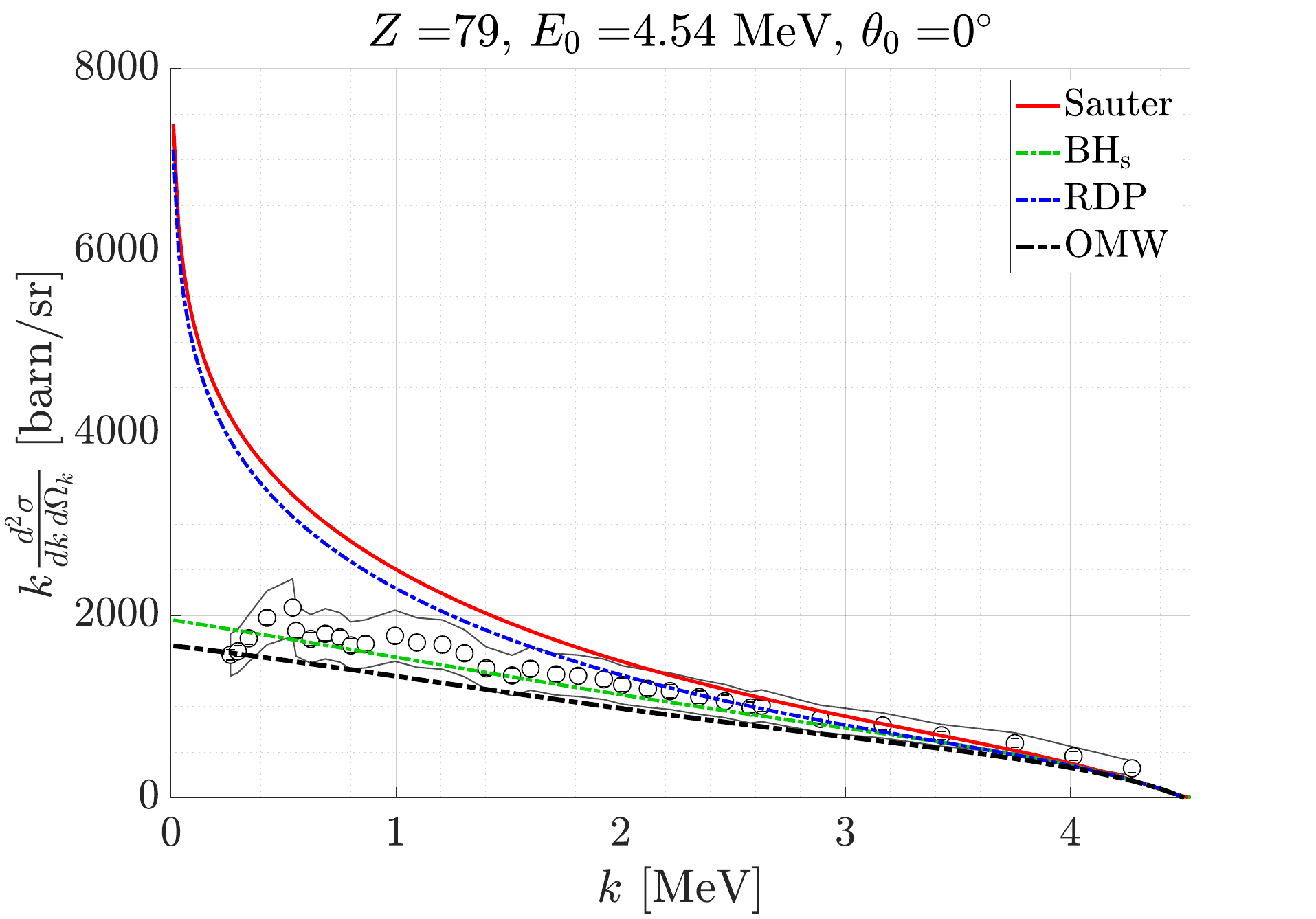}
    \includegraphics[width=0.49\linewidth]{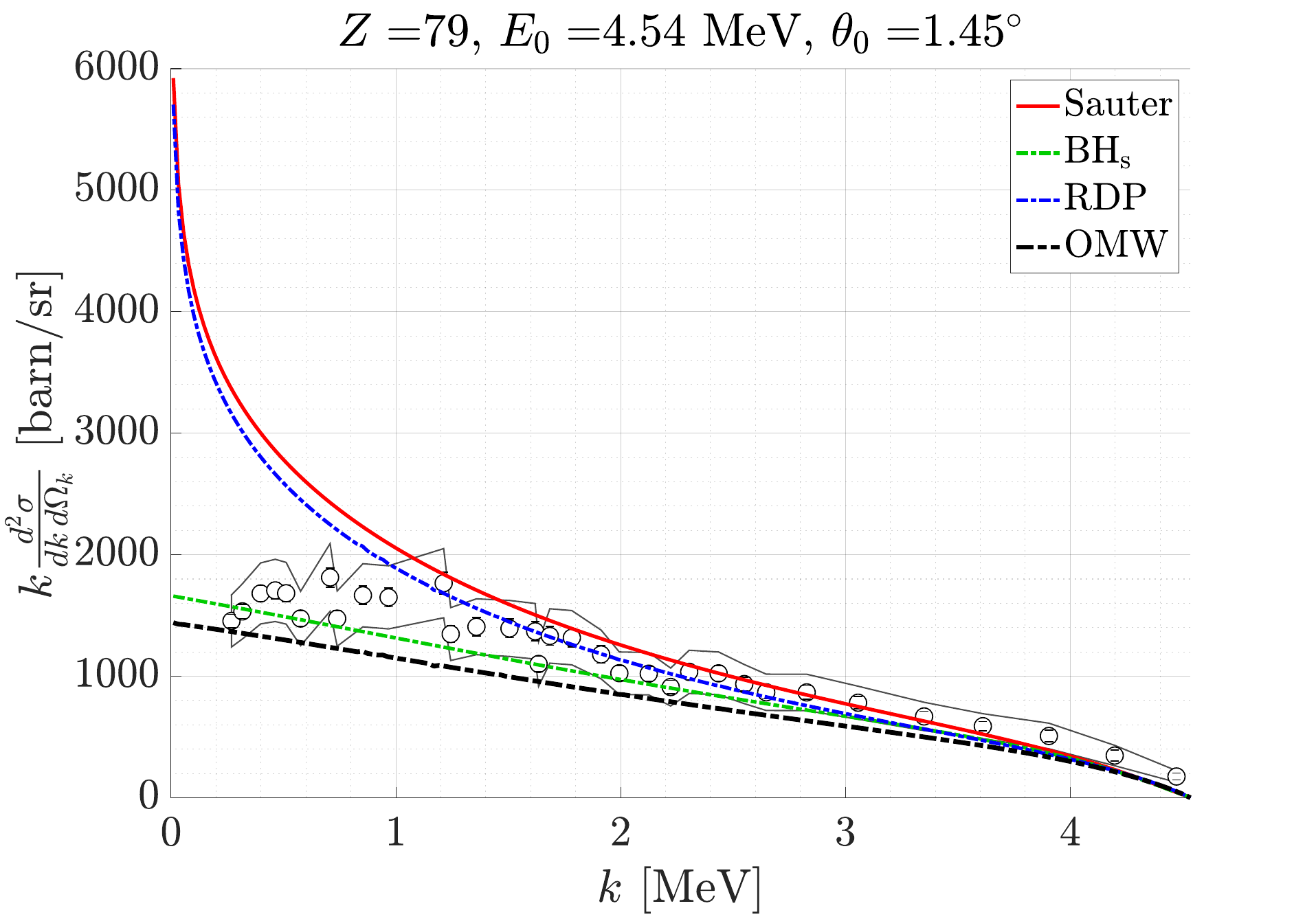}\\[1ex]
    
    \includegraphics[width=0.49\linewidth]{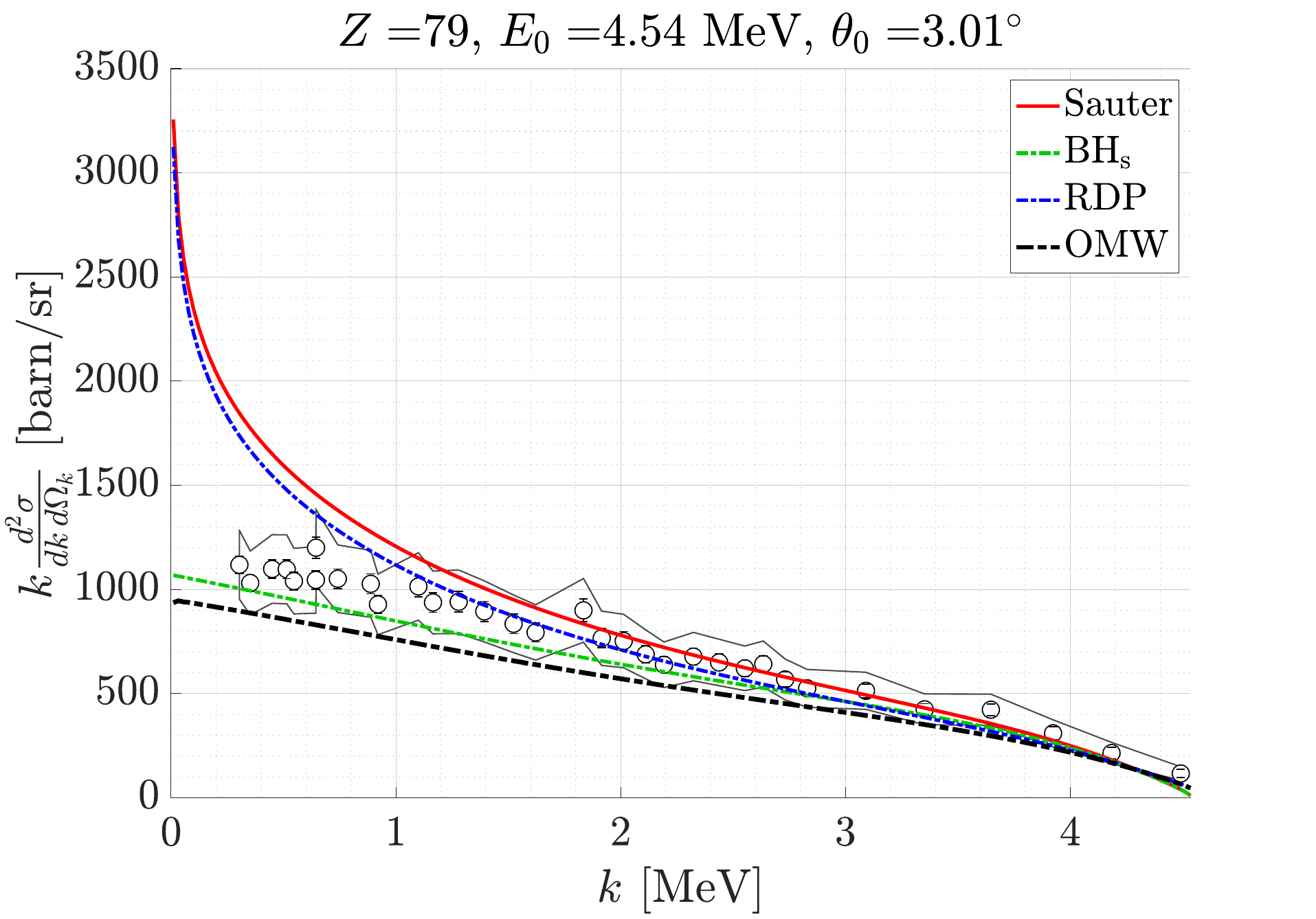}
    \includegraphics[width=0.49\linewidth]{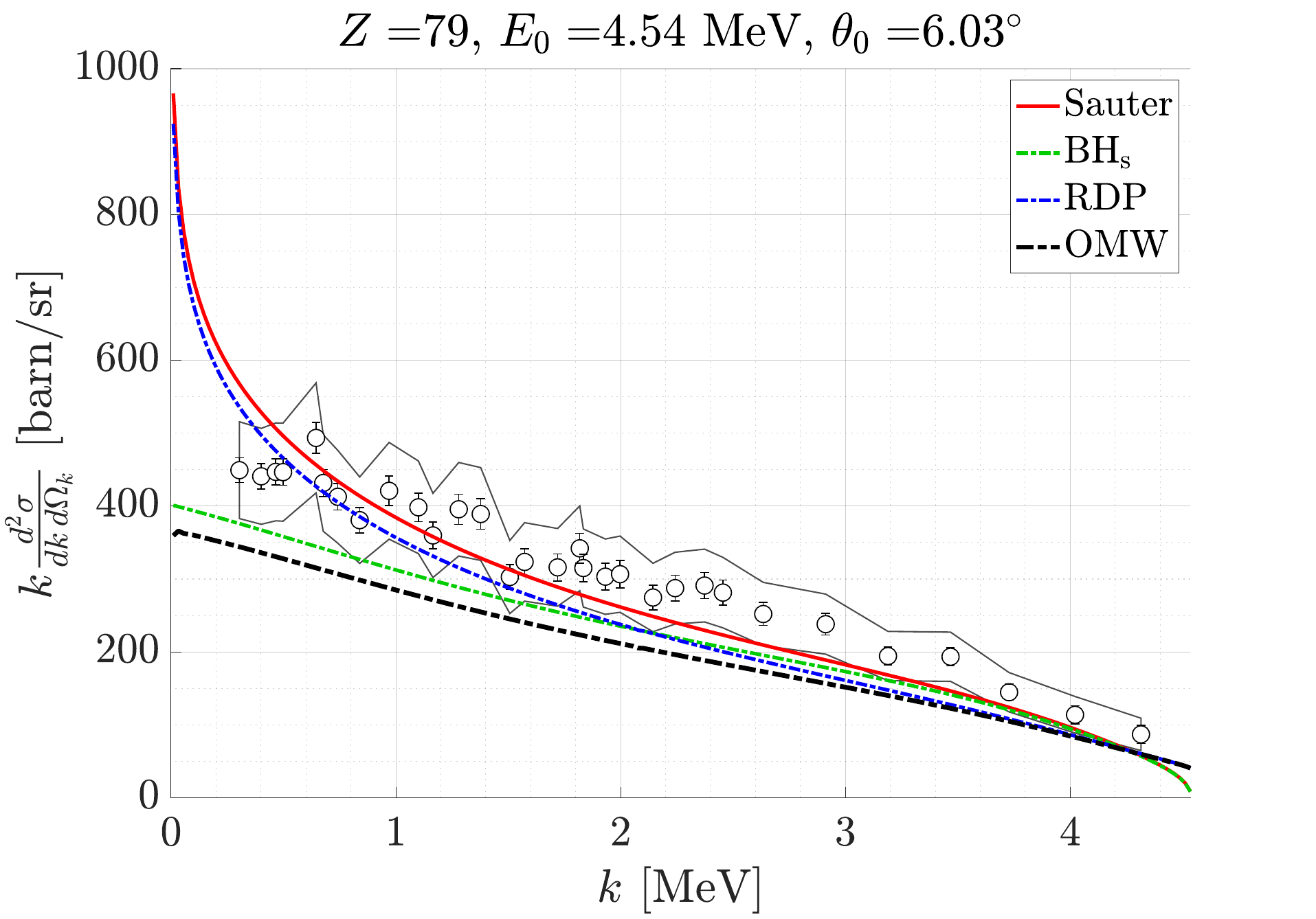}\\[1ex] 
    \caption{Comparison of the theoretical model components against the experimental results by Starfelt and Koch \cite{Starfelt56}, for gold at several low angles where the screening is important. The error bars represent statistical uncertainties, while the contours indicate the sum of the statistical and systematic errors. The incoming electron has an energy of 4.54 MeV.}
    \label{fig:Gold_4_54MeV}
\end{figure*}
\begin{figure*}[!t]
    \centering
    \includegraphics[width=0.49\linewidth]{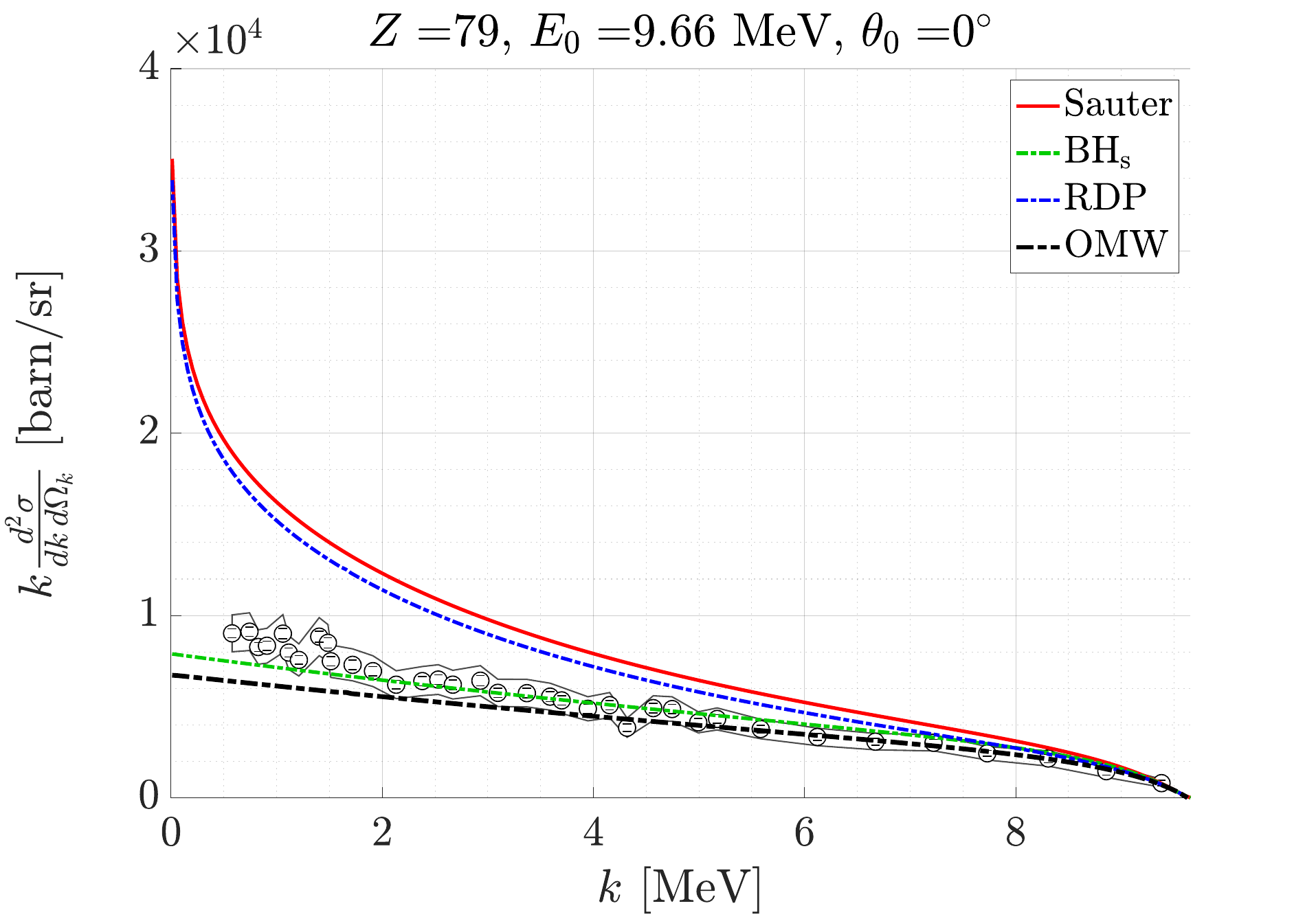}
    \includegraphics[width=0.49\linewidth]{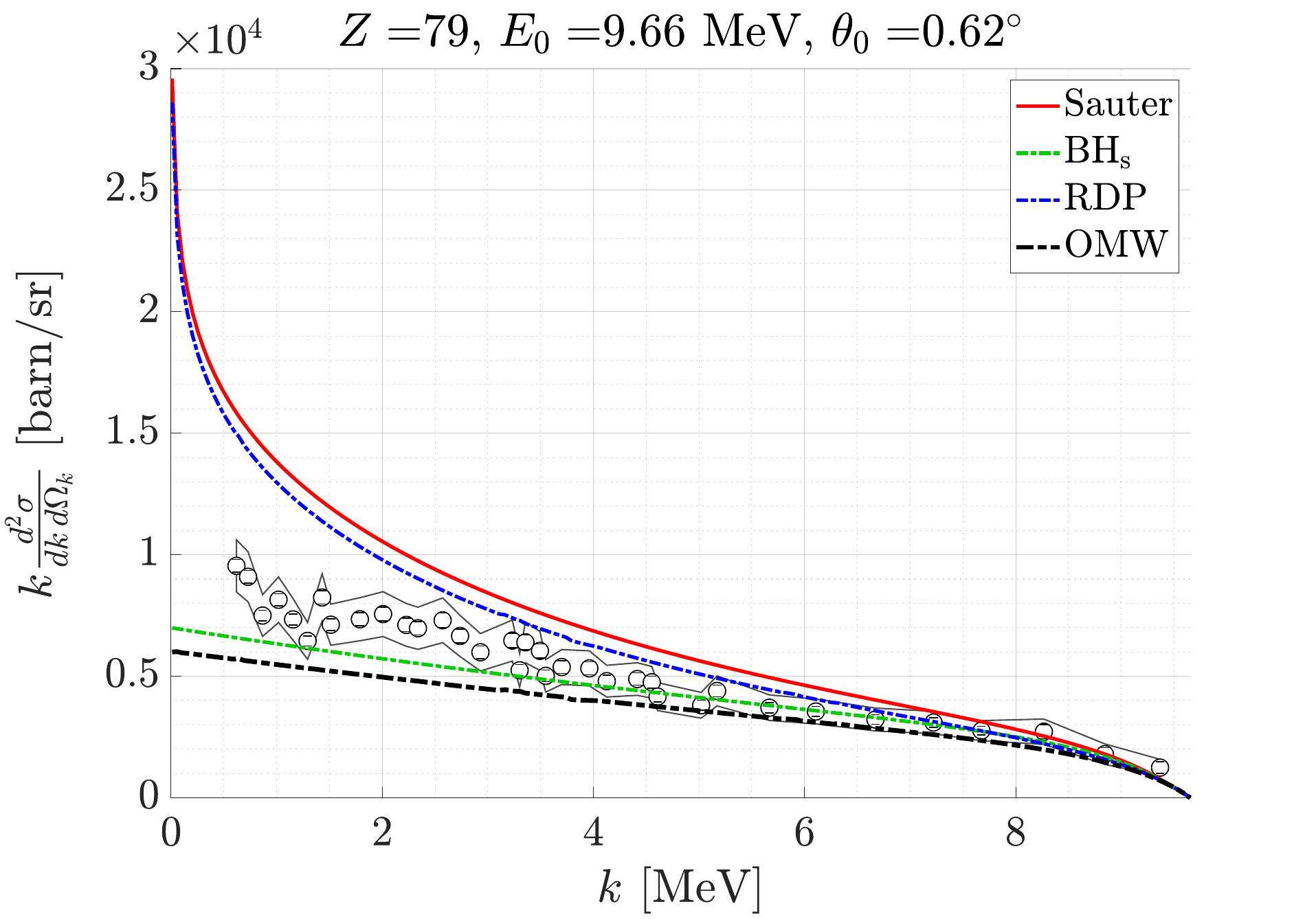}\\[1ex]
    
    \includegraphics[width=0.49\linewidth]{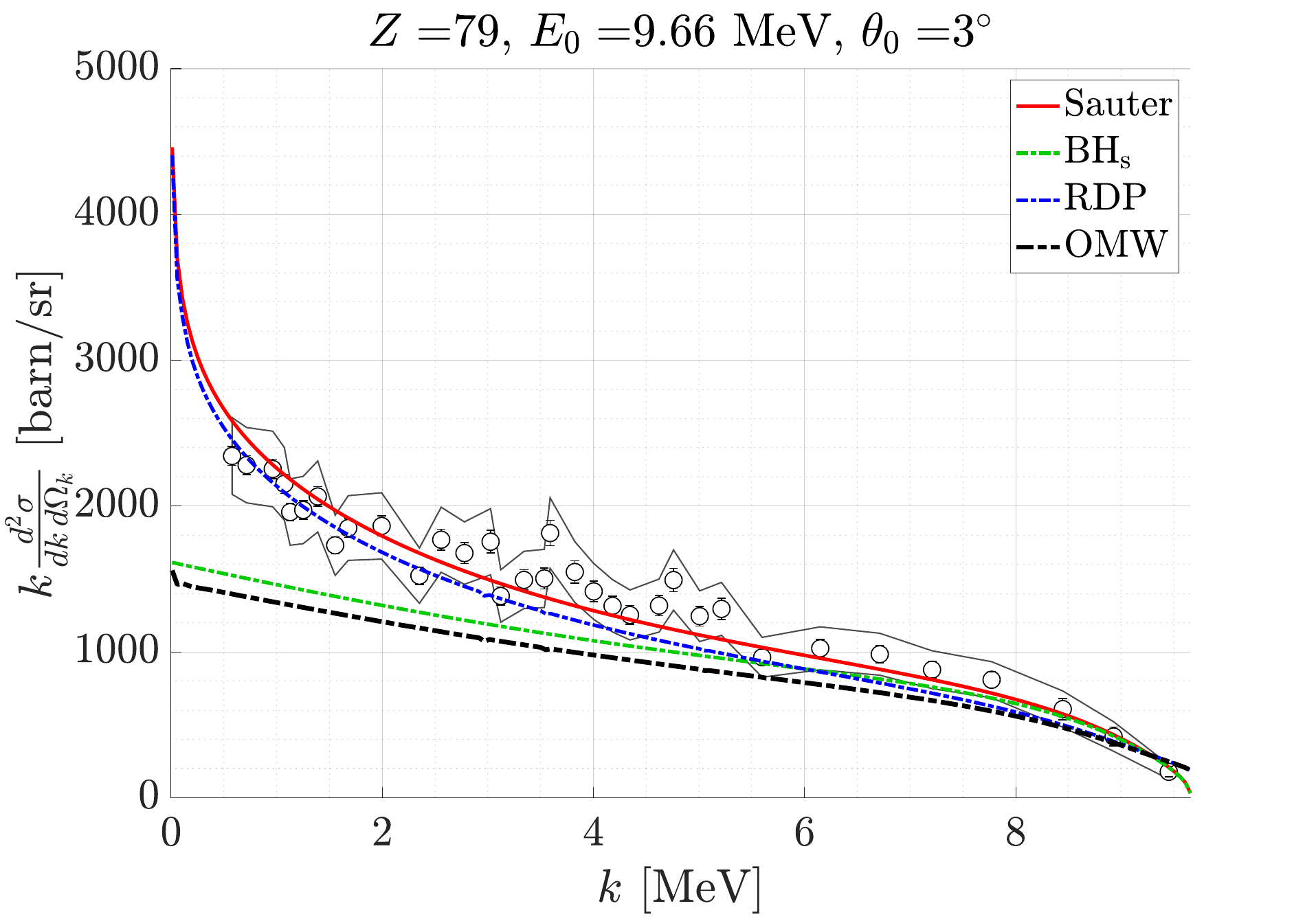}
    \includegraphics[width=0.49\linewidth]{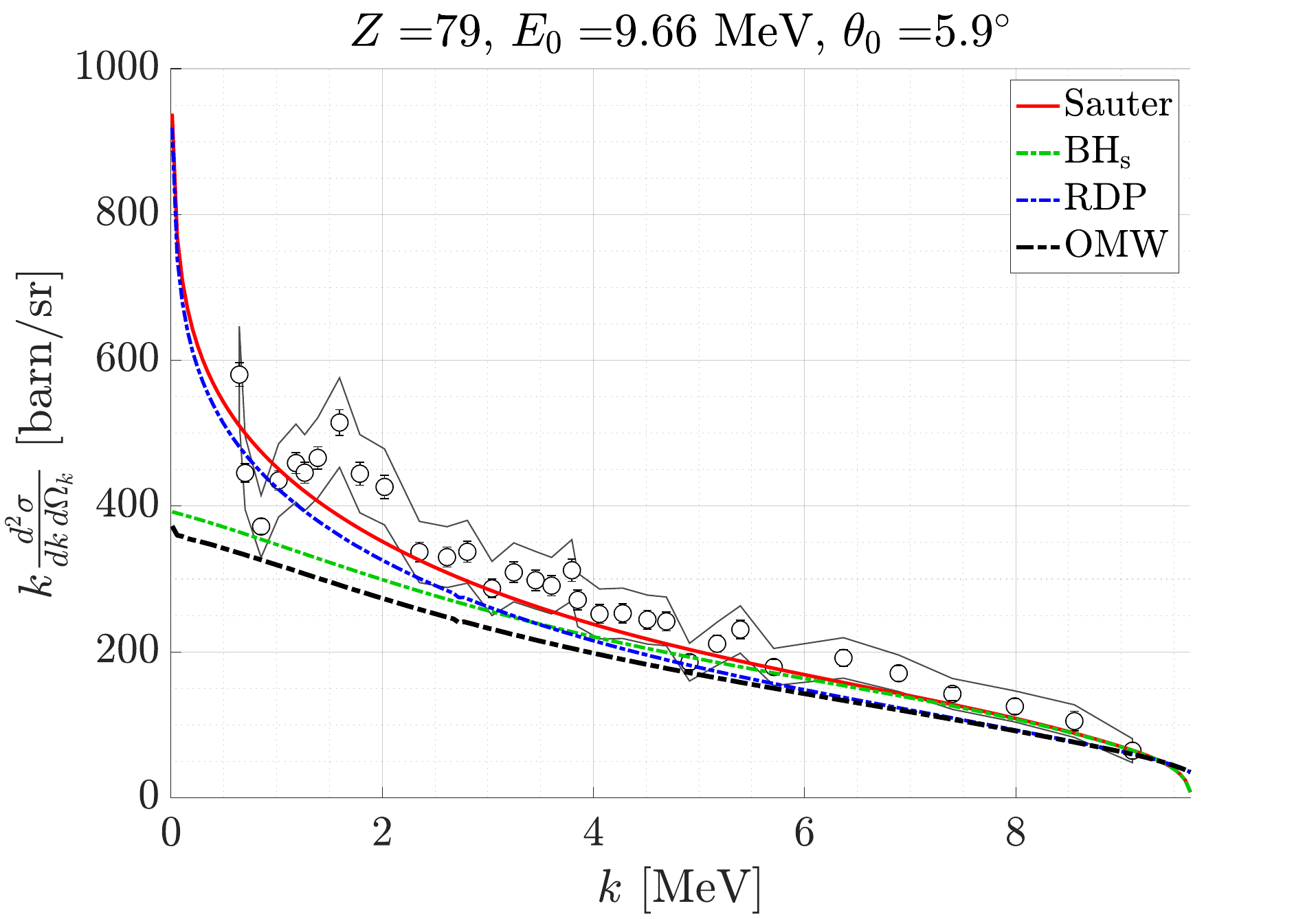}\\[1ex] 
    \caption{Comparison of the theoretical model components against the experimental results by Starfelt and Koch \cite{Starfelt56}, for gold at several low angles where the screening is important. The error bars represent statistical uncertainties, while the contours indicate the sum of the statistical and systematic errors. The incoming electron has an energy of 9.66 MeV.}
    \label{fig:Gold_9_66MeV}
\end{figure*}
The screening model developed in Section~\ref{sec:Screened-Electron-ion-bremsstrah} and analyzed in Section~\ref{sec:Multiyuk} is combined with the exact Coulomb bremsstrahlung correction at the doubly differential level through Eq.~\eqref{eq:OMW}. This formulation enables direct comparison with experimental measurements in the neutral state for arbitrary atomic species, electron and photon energies, and emission angles. 

A substantial amount of experimental data is available in the literature for thin-target bremsstrahlung measurements of the doubly differential cross section (DDCS), covering incident electron energies from a few keV up to about 10 MeV~\cite{Motz55, Starfelt56, MotzPlacidous58, koc59, Rester67, elw69, ResterEdmonson72, LeePratt76}.
In the present study, three representative elements---Al, Sn, and Au---are selected to span low to high atomic numbers and to sample different combinations of incident-electron energy and emission angle. The experimental data of Starfelt and Koch~\cite{Starfelt56} and of Rester and Dance~\cite{Rester67} are employed for direct comparison with the model, in accordance with other recent analyses~\cite{Mangiarotti_2016, MangiarottiMartins_2017, JakubassaMangia_2019}.\footnote{Some results by Rester and Dance were previously obtained by Motz~\cite{Motz55}, but they differ significantly from both the later measurements of Rester and Dance and from theoretical predictions. For this reason, they are not considered in the present work.}

In Figs.~\ref{fig:Z_theta0},~\ref{fig:Gold_4_54MeV},~\ref{fig:Gold_9_66MeV}, the solid red lines represent the Sauter formula, Eq.~\eqref{eq:cross-section=0000202BN-1}; the dashed green curves correspond to the multi-Yukawa Born approximation, Eq.~\eqref{eq:DDCS_MY}; and the dashed blue curves show the integrated RDP result, Eq.~\eqref{eq:RDP_tot}. The total cross section obtained from Eq.~\eqref{eq:OMW} is plotted as a dashed black line. Measurement uncertainties include statistical errors (error bars), which affect both the spectral shape and the absolute magnitude of the cross sections, and systematic errors, which influence only the absolute magnitude~\cite{Starfelt56,Rester67}. The contours represent the combined statistical and systematic uncertainties.

For low atomic numbers, the agreement between the model and experiments is excellent over the entire range of photon energies and emission angles, as shown in the two uppermost plots of Fig.~\ref{fig:Z_theta0}. For higher atomic numbers, the model agrees well with experimental measurements of emission in the forward $\leq10^{\circ}$ direction. However, as illustrated in the lower right-hand panel of Fig.~\ref{fig:Z_theta0} shows that both the Born-approximation cross sections (Sauter and screened Bethe--Heitler) and the FSM-based RDP cross section fail to reproduce the experimental data accurately for high atomic number and large emission angle. This tendency was also noted for tin, although with smaller magnitude. The explanation is a combined breakdown of the Born approximation, as the condition
\allowdisplaybreaks{
\begin{align}
\frac{\alpha Z_s}{\beta_0},\;\frac{\alpha Z_s}{\beta} \ll 1 \label{eq:Born_cond}
\end{align}}
is no longer satisfied, and of the FSM wave-function validity.
The breakdown of the FSM condition is likely the dominant cause, given that Eq.~\eqref{eq:Born_cond} contains no angular dependence, and the lower-left panel ($Z=79$, $\theta_0 = 10^{\circ}$) does not reveal any obvious failure of the Born-based cross sections (Sauter and screened Bethe--Heitler).

At larger angles, where screening effects are weaker, the Sauter and Bethe--Heitler results are expected to approach each other, since the screening correction within the Born approximation becomes small. In this regime, the dominant contribution should therefore arise from the RDP cross section, which lies significantly below the experimental values for $\theta_0=60^{\circ}$, consistent with the breakdown of the FSM approximation in the high-$q$ region.\footnote{Note that Furry--Sommerfeld--Maue condition~\eqref{eq:FSM_cond} involves the local angle $\vartheta_{pr}$ between the electron momentum and its position vector in the Coulomb field, rather than the measured photon emission angle $\theta_0$. Although these two quantities are distinct, they are correlated through the recoil momentum ${\bm{q}}={\bm{p}}_0-{\bm{p}}-{\bm{k}}$ as larger ${q}$ corresponds to smaller effective electron–nucleus separations and therefore to stronger local deflections~\cite{haug2004elementary}. The FSM approximation is expected to deteriorate in regions of large ${q}$, which typically occur for large emission angles.} It is however important to note that the corresponding cross sections are two orders of magnitude lower than those obtained for emissions at $10^{\circ}$.

The onset of this behavior also appears at smaller emission angles for higher incident electron energies, as shown by comparison with data in Figs.~\ref{fig:Gold_4_54MeV} and~\ref{fig:Gold_9_66MeV}.
The validity of the FSM condition~\eqref{eq:FSM_cond} becomes increasingly restricted at higher incident energies, since the momentum transferred to the nucleus, ${q}$, rises for a given emission angle, probing smaller distances from the nucleus according to $r \sim 1/{q}$~\cite{haug2004elementary}.
As the relevant interaction radius decreases, the local deflection $\vartheta_{pr}$ of the electron in the Coulomb field grows, and the FSM approximation can fail even for comparatively small observable angles.
This trend is evident in Fig.~\ref{fig:Gold_4_54MeV} for 4.54~MeV electrons incident on gold ($Z=79$), where a noticeable discrepancy between the model and the experimental measurements already appears at $\theta_{0}\simeq6^\circ$. At such angles, however, the corresponding emission intensity is an order of magnitude lower than in the purely forward direction ($\theta_{0}=0^\circ$).  
It is important to note that in the near-forward direction $(\theta_{0}\sim0^\circ)$, the RDP cross section decreases sharply in the high-frequency limit, in a manner similar to the Born results (see the two upper panels of Fig.~\ref{fig:Gold_4_54MeV}). Although this behavior is mitigated as the emission angle increases, the RDP cross section cannot be expected to be accurate in the tip region, since the next-to-leading-order correction relies on the assumption that the scattered electron remains highly relativistic. By contrast, the leading-order (EH) cross section does not exhibit such a decrease in the tip region, although it is not shown here.
As the emission angle increases, all theoretical components (RDP, Sauter, and BH$_{\mathrm{s}}$) shift downward relative to the experimental data, such that their combination,~Eq.~\eqref{eq:OMW}, lies systematically below the measurements, even though these angles remain smaller than in Fig.~\ref{fig:Z_theta0}.

A similar trend is observed when comparing with the highest available thin-target bremsstrahlung cross sections, measured at an incident electron energy of 9.66~MeV. As shown in Fig.~\ref{fig:Gold_9_66MeV}, the theoretical components exhibit a consistent downward displacement relative to the experimental points.

At these higher energies, an additional source of uncertainty arises from the photomultiplier response function used by Starfelt and Koch, which was obtained through Monte Carlo simulations performed by other authors~\cite{Starfelt56}. Nevertheless, the systematic downward shift of all theoretical components strongly indicates that the principal cause of the discrepancy is the breakdown of the FSM approximation in the high-$q$ regime, where the local perturbative expansion ceases to hold.

\begin{figure}[hbtp!]
\centering
\includegraphics[width=0.49\textwidth]{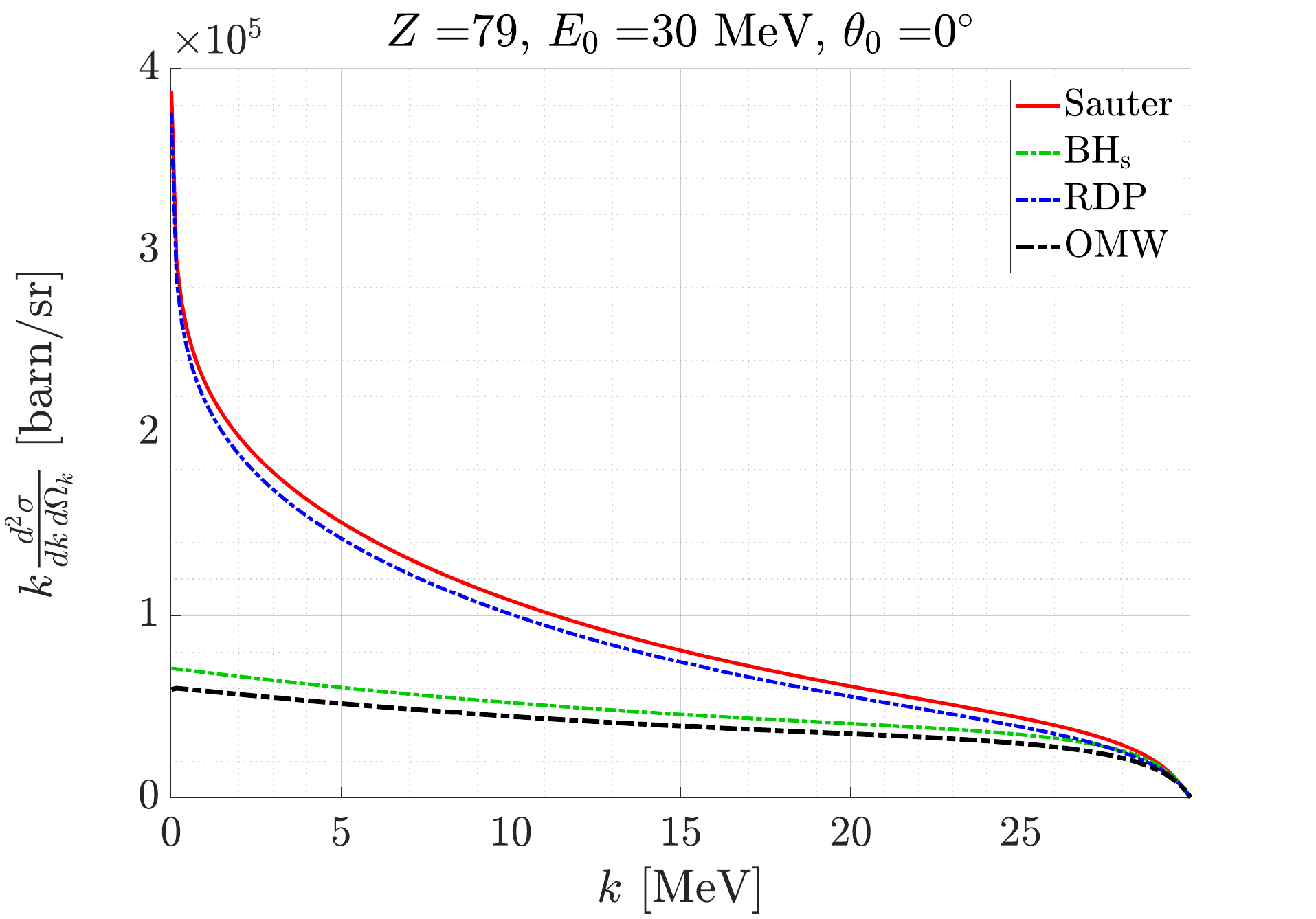}\\
\includegraphics[width=0.49\textwidth]{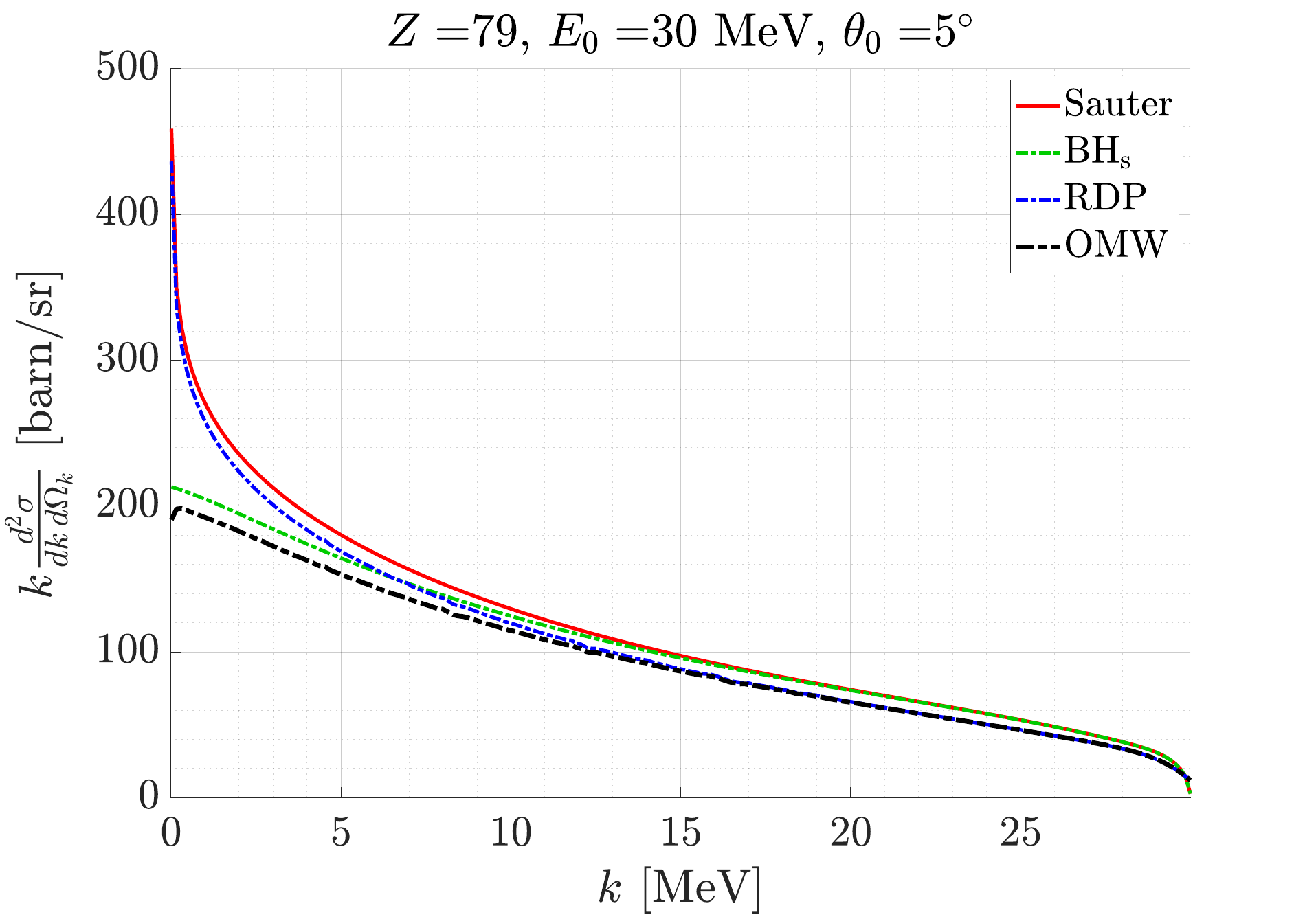}\caption{Top: Theoretical model components for a 30 MeV electron incident on gold, with emission at $\theta_0=0^{\circ}$. Bottom: Identical, for $\theta_0=5^{\circ}$.}
\label{fig:app_higherE_Gold}
\end{figure}

Roche, Ducos, and Proriol originally indicated that their formulation remains applicable up to electron energies of approximately 50~MeV~\cite{roc72}. In the present work, this framework is employed in its consistently ordered form (see Eq.~\eqref{eq:sigma_ei_rdp_2}). Figure~\ref{fig:app_higherE_Gold} shows the components of the model in Eq.~\eqref{eq:OMW} for an incident electron energy of 30~MeV on a gold target. The various contributions behave consistently and exhibit no qualitative discrepancies relative to the trends observed at lower energies. The choice of 30~MeV corresponds to approximately three times the highest incident energy for which experimental data are currently available. These results suggest that the present formulation can reasonably be applied at such energies, provided the system remains in a regime where the breakdown of the FSM approximation is limited. When the FSM approximation is no longer adequate, an exact treatment of the Coulomb field using Dirac wave functions provides a natural alternative, assuming that screening corrections can still be incorporated additively within the OMW framework.
\section{Conclusion and Outlooks}\label{sec:conclusions}
In this paper, the general formalism describing the calculation of bremsstrahlung differential cross sections was reviewed, addressing the main challenges still faced by existing theoretical models. 
The commonly adopted expression for the doubly differential cross section (DDCS) is based on 
the cross section in a pure Coulomb field, calculated within the Furry--Sommerfeld--Maue (FSM) picture and corrected for screening effects in the Born approximation. 
The novelty of the present work lies in a new formulation of the screening correction such that (i) the model accounts for screening in partially ionized states, and (ii) the correction is expressed in fully analytic form, allowing instantaneous evaluation and efficient exploration of large parameter spaces. 
In addition, the Coulomb-field contribution entering this formulation was evaluated using the corrected next-to-leading-order Roche--Ducos--Proriol (RDP) expression.

The partially screened DDCS was obtained by extending the Bethe--Heitler cross section through the inclusion of an atomic form factor representing partially ionized states, described by a multi-Yukawa potential~\cite{sav23}.  Building on existing results for neutral screened bremsstrahlung, an analytical expression was derived for arbitrary ionization levels.

The model was applied to a range of atomic numbers, emission angles, and incident electron energies, while varying internal parameters such as the number of Yukawa exponentials used to represent the screened potential. The results show that the emission intensity does not increase monotonically with ionization: for certain photon energies, a minimum appears at higher ionization levels. This behavior arises from the non-monotonic dependence of the atomic form factors on ionization, which reflects oscillations in the radial bound-electron densities as the ionization state changes.

The performance of the screened model of Eq.~\eqref{eq:OMW} in the neutral case was evaluated by comparison with experimental measurements across a range of atomic numbers, emission angles, and incident electron energies. Excellent agreement was obtained for forward emission angles and incident energies up to about 5~MeV, from low- to high-$Z$ elements. At larger emission angles, the model systematically underestimates the measured spectra, consistent with previous observations~\cite{Mangiarotti_2016, MangiarottiMartins_2017, JakubassaMangia_2019}. This discrepancy is attributed to the breakdown of the FSM approximation at larger deflection angles and becomes apparent at smaller angles as the energy increases. However, this limitation is not critical, since cross sections in these regions are several orders of magnitude smaller than in forward emission. In practical situations, angular integration and the finite spatial and momentum spread of the electron beam further increase the contribution of forward emission.

Addressing the identified limitations may require moving beyond the present set of approximations and solving the Dirac equation for exact electron wave functions in the screened potential of ionized atoms. Such an extension would be computationally demanding but represents a natural direction for future work. Additionally, new and extended experimental data would be highly valuable for further benchmarking and refinement of the models.

\section*{Data Availability Statement}
The data that support the findings of this article are openly available \cite{brem_cross_sec_release}.

\section*{Acknowledgments}
The authors are indebted to A.~Mangiarotti for providing some of the tabulated experimental data, and thankful to A.~Mangiarotti and K.~Barbey for valuable discussions.

\section*{Author Contributions}
\textbf{S.~Guinchard:} Conceptualization (equal); Formal analysis (equal); Methodology (lead); Software (equal); Writing –
original draft (lead); Writing – review \& editing (lead). \textbf{Y.~Savoye-Peysson:} Conceptualization (equal); Formal analysis (equal); Software (equal); Writing – review \& editing (equal). \textbf{J.~Decker:} Formal analysis (equal); Software (equal); Writing – review \& editing (equal).

\FloatBarrier
\bibliographystyle{alpha}
\bibliography{bibliography}

\end{document}